\documentclass[
  journal=jacsat,
  manuscript=article,
    layout=twocolumn
]{achemso}

\usepackage[version=3]{mhchem} 

\usepackage{booktabs}
\usepackage{siunitx}

\usepackage{titlesec}

\titleformat{\section}
  {\large\bfseries}
  {}      
  {0pt}
  {}

\titlespacing*{\section}
  {0pt}
  {1.2ex} 
  {0.8ex} 

\titleformat{\subsection}
  {\normalsize\bfseries} 
  {}      
  {0pt}
  {}

\titlespacing*{\subsection}
  {0pt}
  {1.0ex}
  {0.6ex}

\titleformat{\paragraph}
  {\footnotesize\bfseries}
  {}      
  {0pt}
  {}

\titlespacing*{\paragraph}
  {0pt}
  {0.6ex}
  {0.3ex}

\author{Chengchun Liu}
\affiliation{\textit{School of AI for Science, Peking University Shenzhen Graduate School, Shenzhen, 518055, China}}

\alsoaffiliation[Peking University]
{AI for Science (AI4S)-Preferred Program, Peking University Shenzhen Graduate School, Shenzhen 518055, China}
\alsoaffiliation
{State Key Laboratory of Advanced Waterproof Materials, School of Materials Science and Engineering, Peking University, Beijing 100871, China}

\author{Wendi Cai}
\affiliation{\textit{School of AI for Science, Peking University Shenzhen Graduate School, Shenzhen, 518055, China}}

\alsoaffiliation[Peking University]
{AI for Science (AI4S)-Preferred Program, Peking University Shenzhen Graduate School, Shenzhen 518055, China}

\author{Boxuan Zhao}
\affiliation{\textit{School of AI for Science, Peking University Shenzhen Graduate School, Shenzhen, 518055, China}}

\alsoaffiliation[Peking University]
{AI for Science (AI4S)-Preferred Program, Peking University Shenzhen Graduate School, Shenzhen 518055, China}
\alsoaffiliation
{State Key Laboratory of Advanced Waterproof Materials, School of Materials Science and Engineering, Peking University, Beijing 100871, China}

\author{Fanyang Mo}
\email{fmo@pku.edu.cn}
\affiliation{\textit{School of AI for Science, Peking University Shenzhen Graduate School, Shenzhen, 518055, China}}

\alsoaffiliation[Peking University]
{AI for Science (AI4S)-Preferred Program, Peking University Shenzhen Graduate School, Shenzhen 518055, China}
\alsoaffiliation
{State Key Laboratory of Advanced Waterproof Materials, School of Materials Science and Engineering, Peking University, Beijing 100871, China}

\alsoaffiliation{School of Advanced Materials, Peking University Shenzhen Graduate School, Shenzhen 518055, China}
\alsoaffiliation{Guangdong Provincial Key Laboratory of Nano-Micro Materials Research, Peking University Shenzhen Graduate School, Shenzhen 518055, China}

\title[An \textsf{achemso} demo]
  {A Cross-Domain Graph Learning Protocol for Single-Step Molecular Geometry Refinement}

\begin{document}







\begin{abstract}
Accurate molecular geometries are a prerequisite for reliable quantum-chemical predictions, yet density functional theory (DFT) optimization remains a major bottleneck for high-throughput molecular screening. 
Here we present GeoOpt-Net, a multi-branch SE(3)-equivariant geometry refinement network that predicts DFT-quality structures at the B3LYP/TZVP level of theory in a single forward pass starting from inexpensive initial conformers generated at a low-cost force-field level.
GeoOpt-Net is trained using a two-stage strategy in which a broadly pretrained geometric representation is subsequently fine-tuned to approach B3LYP/TZVP-level accuracy, with theory- and basis-set-aware calibration enabled by a fidelity-aware feature modulation (FAFM) mechanism.
Benchmarking against representative approaches spanning classical conformer generation (RDKit), semiempirical quantum methods (xTB), data-driven geometry refinement pipelines (Auto3D), and machine-learning interatomic potentials (UMA) on external drug-like molecules from the ZINC20 database demonstrates that GeoOpt-Net achieves sub-milli-\AA{} all-atom RMSD with near-zero B3LYP/TZVP single-point energy deviations, indicating DFT-ready geometries that closely reproduce both structural and energetic references.
Beyond geometric metrics, GeoOpt-Net generates initial guesses intrinsically compatible with DFT convergence criteria, yielding nonzero ``All-YES'' convergence rates (65.0\% under loose and 33.4\% under default thresholds) while all baselines achieve 0\%, and substantially reducing re-optimization steps and wall-clock time. 
GeoOpt-Net further exhibits smooth and predictable energy scaling with molecular complexity while preserving key electronic observables such as dipole moments. Collectively, these results establish GeoOpt-Net as a scalable, physically consistent geometry refinement framework that enables efficient acceleration of DFT-based quantum-chemical workflows.
\end{abstract}

\section{Introduction}

\begin{figure*}[ht]
    \centering
    \includegraphics[width=0.95\textwidth]{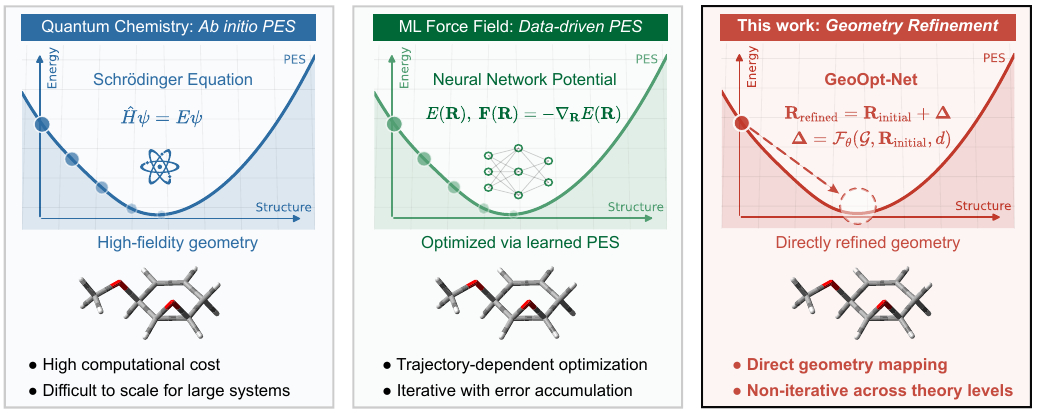}
    \caption{
    Conceptual comparison of geometry optimization paradigms.
    Left: conventional quantum-chemical optimization on an \textit{ab initio} potential energy surface.
    Middle: iterative relaxation driven by a learned machine-learning force field.
    Right: GeoOpt-Net directly refines molecular geometries in a single forward pass without explicit
    potential energy surface construction or iterative optimization.
    }
    \label{fig1}
\end{figure*}

Accurate three‐dimensional molecular geometries are the bedrock of computational chemistry~\cite{gillespie2005models,sadowski1993atoms}, underpinning reliable predictions of reaction pathways~\cite{bruice2006computational,van20243dreact}, spectroscopic signatures~\cite{han2024accurate,liu2024infrared} and the rational design of catalysts~\cite{vogiatzis2018computational}, functional materials~\cite{fang2022geometry} and drug candidates~\cite{shen2021out,blundell2002high}.  At the heart of these applications lies the exploration of the potential energy surface (PES), a high‐dimensional hypersurface defined by the Born–Oppenheimer approximation, whose minima correspond to stable conformers and whose saddle points mark transition states~\cite{manzhos2020neural}.  Quantum‐mechanical (QM) methods—from Hartree–Fock (HF)~\cite{slater1951simplification,fischer1977hartree} through Kohn–Sham density functional theory (DFT)~\cite{kohn1996density}—evaluate energies and forces via the self‐consistent solution of the electronic Schrödinger equation, with computational cost scaling steeply~\cite{riley2010stabilization} (approximately \(O(N^3)\)–\(O(N^4)\)) in the number of basis functions \(N\) (Fig.~\ref{fig1} left). In practice, hybrid density functional methods such as B3LYP combined with triple-$\zeta$ basis sets (e.g., TZVP) are widely adopted as a pragmatic reference for molecular geometries~\cite{becke1993density,lee1988development,weigend2005balanced}. Numerous benchmarks have shown that B3LYP/TZVP geometries are close to basis-set–converged structures for typical organic molecules, with residual geometric differences having only minor impact on relative energies, spectroscopic properties, and chemical trends~\cite{goerigk2011thorough,mardirossian2017thirty}. As a result, B3LYP/TZVP has become a de facto standard for geometry optimization in large-scale and high-throughput quantum-chemical workflows.

Recent advances in machine learning (ML) have introduced multiple strategies to accelerate molecular geometry prediction.
One direction focuses on learning differentiable surrogate models of the potential energy surface~\cite{schutt2017schnet,unke2021machine,unke2021spookynet}, allowing molecular geometries to be optimized using ML-derived forces.
While these models significantly reduce the cost of individual energy or gradient evaluations, they still rely on iterative optimization loops and may suffer from limited transferability when applied to molecules outside the training distribution (Fig.~\ref{fig1} middle).
Recent methods such as UMA~\cite{wood2025family} further advance this line of work by training machine-learned potentials on hybrid density functionals, thereby improving the fidelity of PES-based geometry optimization, though still within the iterative force-driven paradigm.
An alternative direction bypasses local optimization altogether by directly generating 3D coordinates from molecular graphs in a single inference step. 
Global search approaches such as ConfGF~\cite{shi2021learning},
DMCG~\cite{zhu2022direct}, and GeoDiff~\cite{xu2022geodiff} attempt to model the full conformational distribution using score-matching, diffusion, or flow-based generative processes.
Although they eliminate explicit energy evaluations during inference, these methods typically require multiple sampling steps per molecule and often produce geometries that fail to satisfy standard DFT convergence criteria, thereby necessitating downstream quantum refinement (see Fig.~S1 in the SI for detailed results).
By contrast, local refinement strategies such as Auto3D (ANI-2x, AIMNET)~\cite{liu2022auto3d} and RDKit’s ETKDG sampler~\cite{riniker2015better} typically generate conformers only at the force-field (FF) accuracy level. While these methods provide useful initial guesses that are closer to equilibrium structures than random geometries, they still rely on subsequent quantum-mechanical optimization to achieve DFT-level accuracy.

Beyond purely ML-based approaches, semiempirical quantum methods such as
GFN-xTB~\cite{bannwarth2019gfn2} offer a pragmatic compromise
between DFT-level accuracy and computational efficiency. xTB achieves
significantly lower cost by parametrizing tight-binding Hamiltonians, and is
widely used for large-scale conformer sampling and as a pre-optimization step
prior to higher-level DFT refinement. However, despite their speed advantage,
xTB optimizations still require iterative gradient evaluations and may suffer
from systematic deviations in bond lengths, angles, and dihedrals relative to
DFT, particularly for charged or electronically complex systems. These
limitations highlight the broader need for approaches that can deliver
near-DFT accuracy with the efficiency of a single inference step.

Beyond these generative models, several methods incorporate geometric constraints to guide structure prediction. GraphDG~\cite{Simm2020GraphDG} infers probabilistic distance matrices and reconstructs conformers via classical distance geometry. CGCF~\cite{xu2021learning} employs coarse-to-fine message passing conditioned on graph context to iteratively refine geometries. GeoMol~\cite{ganea2021geomol} predicts torsional angles and builds 3D coordinates sequentially from molecular trees, enabling efficient structure generation but without enforcing SE(3)-equivariance or leveraging vector-valued representations.

Related developments can also be found in the generation of transition state (TS) geometries, which share the challenge of recovering chemically consistent 3D structures from sparse graph-level information. TSNet~\cite{jackson2021tsnet} predicts TS structures by jointly encoding reactants and products through a dual-graph architecture. TSDiff~\cite{kim2024diffusion} adapts diffusion-based generation to the TS setting, while React-OT~\cite{duan2025optimal} formulates structure generation as an optimal transport problem over the PES. Although these models target reaction pathways rather than equilibrium states, they adopt similar principles of geometric inductive bias, graph-to-structure translation, and symmetry awareness.

\begin{figure*}[t]
    \centering
    \includegraphics[width=0.845\textwidth]{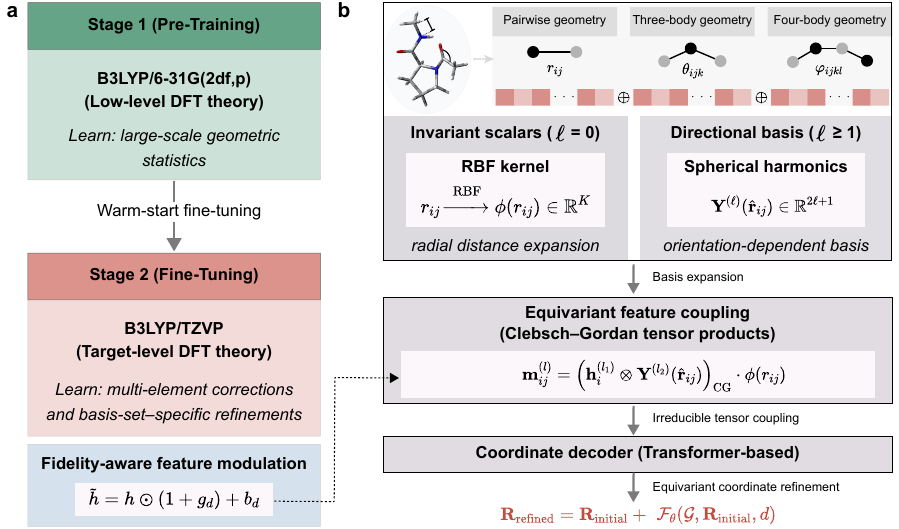}
    \caption{
    \textbf{Two-stage training and SE(3)-equivariant geometric framework of GeoOpt-Net.}
    \textbf{(a)} Two-stage training across quantum-chemical fidelity levels.
    The model is pre-trained on large-scale geometries at the B3LYP/6-31G(2df,p) level and subsequently fine-tuned on target-level B3LYP/TZVP data using a warm-start strategy with fidelity-aware feature modulation (FAFM) mechanism.
    \textbf{(b)} SE(3)-equivariant geometric encoding and coordinate refinement.
    Pairwise distances ($r_{ij}$), bond angles ($\theta_{ijk}$), and dihedral angles ($\varphi_{ijkl}$) are encoded as invariant scalar features ($\ell=0$) via radial basis functions and as directional features ($\ell\ge1$) via spherical harmonics.
    Equivariant features are coupled through Clebsch--Gordan tensor products and decoded by a Transformer-based module to produce SE(3)-equivariant coordinate updates.
    }
    \label{fig2}
\end{figure*}

Geometric deep learning addresses these challenges by embedding physical equivariance directly into neural architectures~\cite{bronstein2017geometric,atz2021geometric}. SE(3)-equivariant graph networks represent atomic features as tensor fields that transform consistently under rotations and translations~\cite{fuchs2020se,batatia2025design}. Messages integrate learnable radial basis expansions of interatomic distances with spherical harmonics of angular coordinates via Clebsch--Gordan tensor products~\cite{andersen1967clebsch}, ensuring each layer’s outputs respect rigid motions and capture local bond geometry, angular dependencies, and global topology.

Invariant directional message‐passing networks such as DimeNet~\cite{gasteiger2020directional} and GemNet~\cite{gasteiger2021gemnet} explicitly incorporate bond angles and torsional dihedrals via designed angular features. Equivariant frameworks like SE(3)-Transformer~\cite{fuchs2020se} and PaiNN~\cite{schutt2021equivariant} guarantee exact SE(3)-equivariance by decoupling scalar and vector feature channels—applying non‐linear activations only to scalars while updating vectors with learned linear maps. More recent methods, such as LEFTNet~\cite{du2023new} implicitly learn angular and higher‐order geometric information through coordinate‐based MLP embeddings. Despite these advances, few approaches systematically decouple bond, angle, and dihedral streams into separate equivariant branches while retaining a one‐shot inference paradigm.

Here, we introduce \textbf{GeoOpt-Net}, a multi-scale SE(3)-equivariant geometry refinement framework designed to directly bridge low-cost initial conformers and high-level DFT accuracy at the B3LYP/TZVP level of theory. Unlike PES-based machine-learning potentials that rely on iterative force-driven optimization, or global generative models that sample broad conformational distributions, GeoOpt-Net targets the specific task of \emph{single-shot refinement} toward DFT-converged geometries (Fig.~\ref{fig1} right).

GeoOpt-Net explicitly decouples bond lengths, bond angles, and dihedral torsions into three dedicated SE(3)-equivariant graph streams, enabling chemically interpretable and physically consistent geometric representations across local and nonlocal length scales.
These multi-branch embeddings are fused through a lightweight Transformer decoder to predict refined atomic coordinates in a single forward pass. To account for systematic differences across electronic structure methods, GeoOpt-Net is trained using a two-stage, multi-fidelity strategy, in which broadly transferable geometric priors are first learned from large-scale pretraining and subsequently fine-tuned to higher levels of theory via a fidelity-aware feature modulation (FAFM) mechanism.

By combining explicit geometric inductive biases, SE(3)-equivariance, and theory-aware fine-tuning, GeoOpt-Net addresses a critical gap in current molecular modeling pipelines: the absence of a fast, robust, and physically grounded route to DFT-ready geometries at the B3LYP/TZVP level, a widely adopted accuracy–cost balance for organic and drug-like molecules, without iterative quantum-mechanical optimization. Beyond accuracy gains, GeoOpt-Net reshapes the role of machine learning in quantum chemistry by transforming geometry optimization from a costly iterative procedure into a deterministic, single-step refinement, enabling scalable and reproducible quantum-chemical workflows targeting DFT-level accuracy for large molecular systems and high-throughput applications.

\begin{figure*}[t]
  \centering
  \includegraphics[width=1\textwidth]{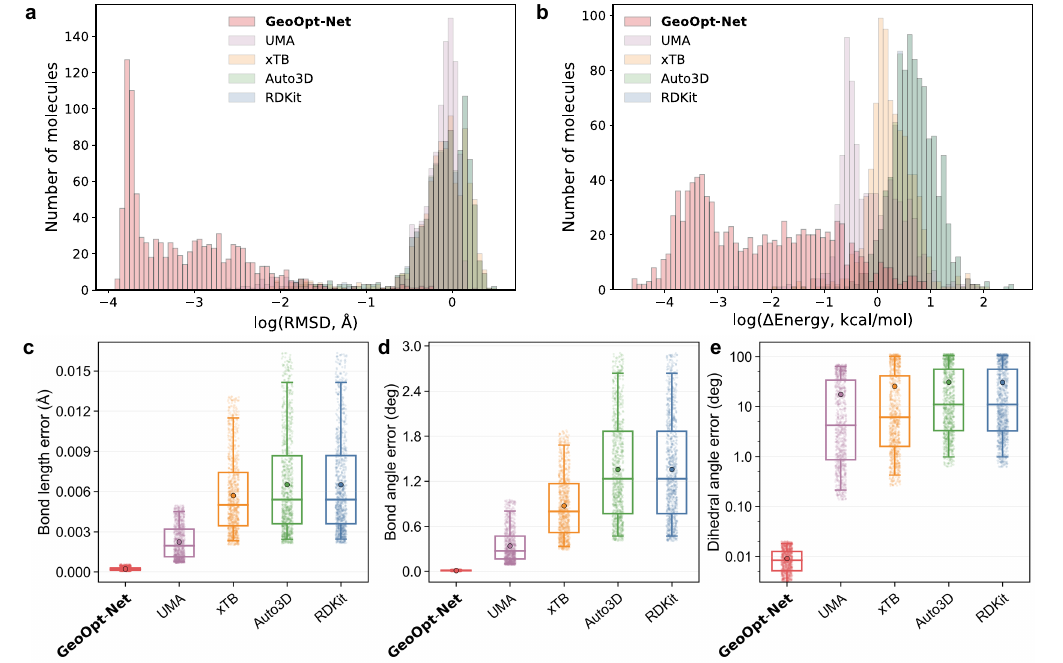}
    \caption{
    \textbf{Benchmarking geometric and energetic accuracy against B3LYP/TZVP references.}
    (a) Logarithmic distribution of all-atom RMSD values relative to B3LYP/TZVP-optimized
    geometries.
    (b) Logarithmic distribution of B3LYP/TZVP single-point energy deviations ($\Delta E$).
    (c--e) Decomposition of geometric errors into (c) bond length, (d) bond angle, and (e) dihedral
    angle deviations, evaluated with respect to the B3LYP/TZVP reference structures.
    }
    \label{fig:geo_benchmark}
\end{figure*}

\section{Datasets and Computational Details}

\subsection{Datasets}

GeoOpt-Net was trained, fine-tuned, and evaluated using multiple quantum-chemical datasets spanning increasing molecular size, elemental diversity, and theoretical levels.

\paragraph{Pre-training datasets.}
The initial pre-training of GeoOpt-Net was performed jointly on the QM9 and QM40 datasets.
QM9~\cite{ramakrishnan2014quantum} comprises approximately 130k small organic molecules containing up to nine heavy atoms (C, N, O, and F), while QM40~\cite{madushanka2024qm40} extends the chemical space to larger and more drug-like systems, covering 160k molecules with elements commonly found in pharmaceutical compounds (C, N, O, S, F, and Cl), representing approximately 88\% of the FDA-approved drug chemical space.
Together, these two datasets provide a combined training corpus of approximately 290k molecules spanning a broad range of molecular sizes and compositions.
All molecular geometries and associated quantum-mechanical properties in both QM9 and QM40 were computed consistently at the B3LYP/6-31G(2df,p) level of theory, ensuring a uniform theoretical framework during the pre-training stage.

\paragraph{Fine-tuning dataset.}
Model fine-tuning was performed using the QMe14S dataset, which consists of 180k small organic molecules featuring substantially increased elemental and functional-group diversity.
QMe14S~\cite{yuan2025qme14s} includes 14 elements (H, B, C, N, O, F, Al, Si, P, S, Cl, As, Se, and Br) and spans 47 functional groups.
All molecular geometries in QMe14S were obtained from density functional theory calculations at the B3LYP/TZVP level, enabling GeoOpt-Net to adapt to a higher basis-set accuracy and broader chemical environments.

\paragraph{External validation dataset.}
To further assess extrapolation performance beyond both the training and fine-tuning domains, an additional validation set was constructed from the ZINC20 database~\cite{irwin2020zinc20}.
From the $\sim$1B ``ZINC20'' collection, 1,000 molecules were selected based on combined constraints on heavy-atom count and rotatable-bond count, ensuring coverage of larger and more conformationally flexible compounds not present in the QM9, QM40, or QMe14S datasets.
These molecules provide a stringent test of GeoOpt-Net under realistic, drug-like structural complexity.

Detailed statistics of dataset splits, along with distributions of heavy-atom counts, rotatable-bond counts, and elemental compositions, are reported in Figures~S2--S4 of the Supporting Information.

\begin{table*}[t]
\centering
\caption{Representative examples comparing optimized molecular geometries.
RMSD values are computed with respect to the DFT reference structures after Kabsch alignment,
and $\Delta E$ denotes the single-point energy difference at the B3LYP/TZVP level of theory.}
\label{tab:three_example_compact}

\setlength{\tabcolsep}{4.5pt}
\renewcommand{\arraystretch}{1.15}

\begin{tabular}{cccccc}
\toprule
Ref.\ (DFT) & \textbf{GeoOpt-Net} & UMA & xTB & Auto3D & RDKit \\
\midrule

\includegraphics[width=0.115\textwidth]{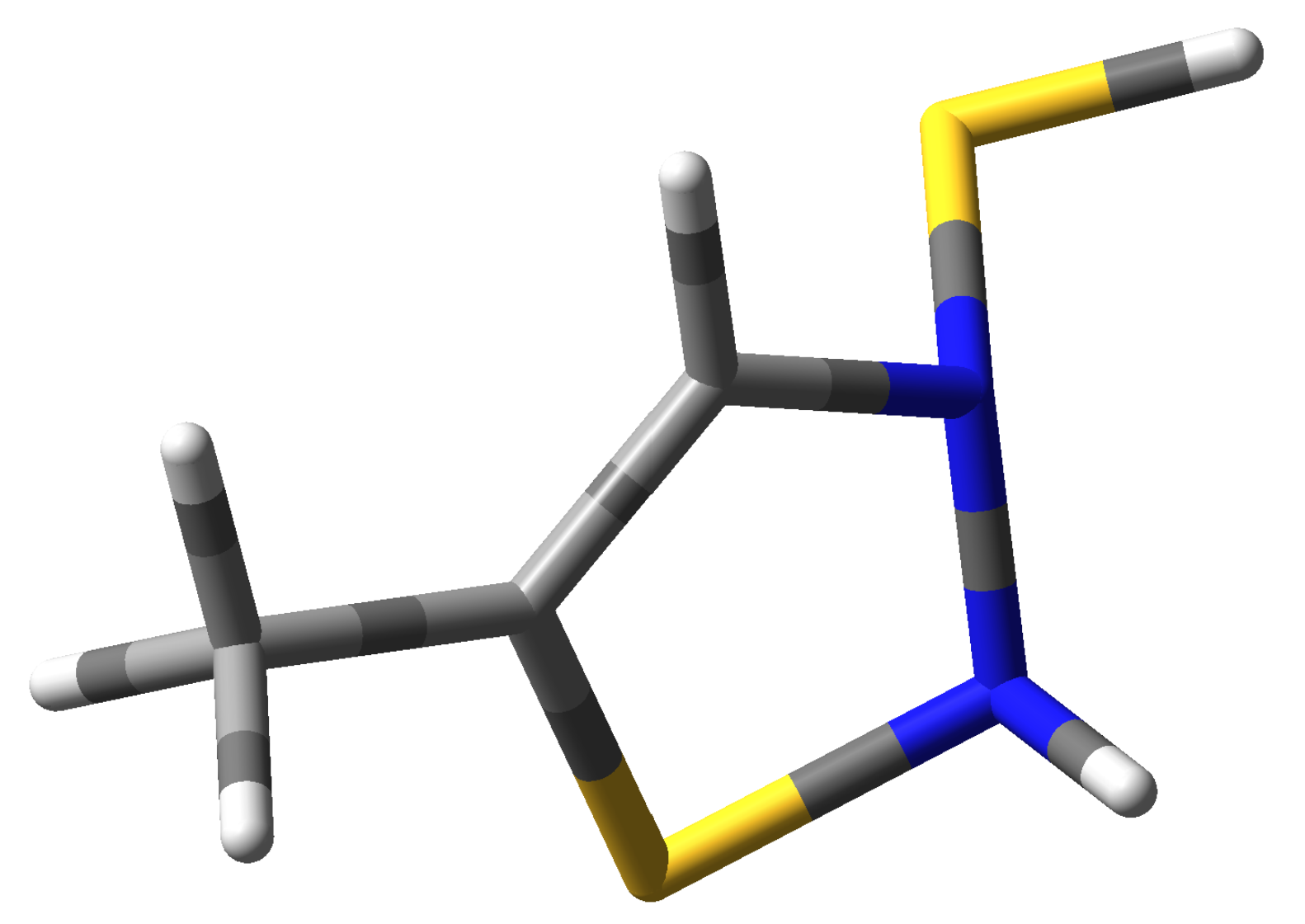} &
\includegraphics[width=0.115\textwidth]{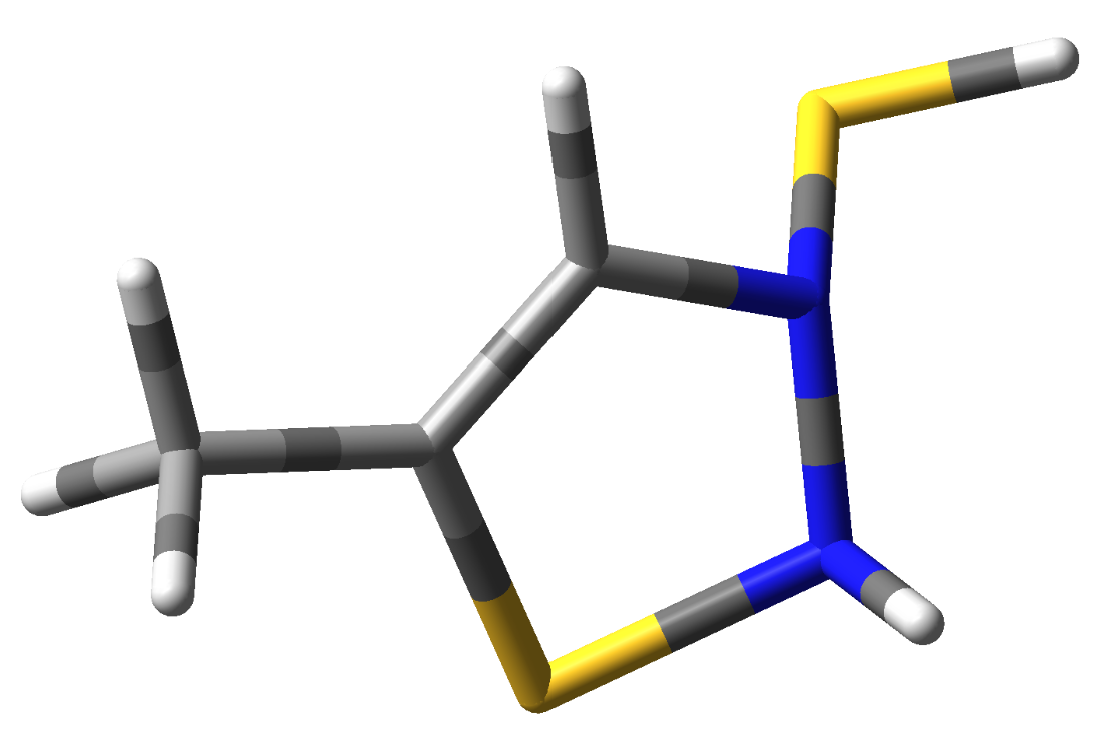} &
\includegraphics[width=0.105\textwidth]{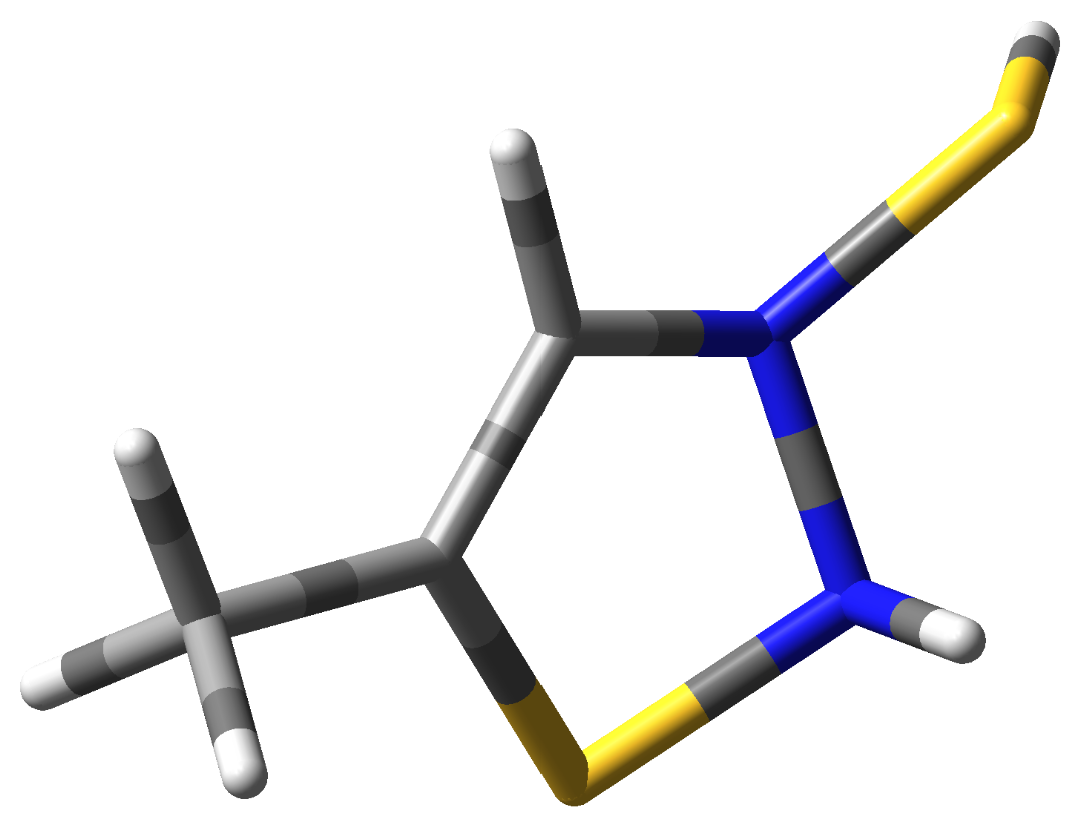} &
\includegraphics[width=0.125\textwidth]{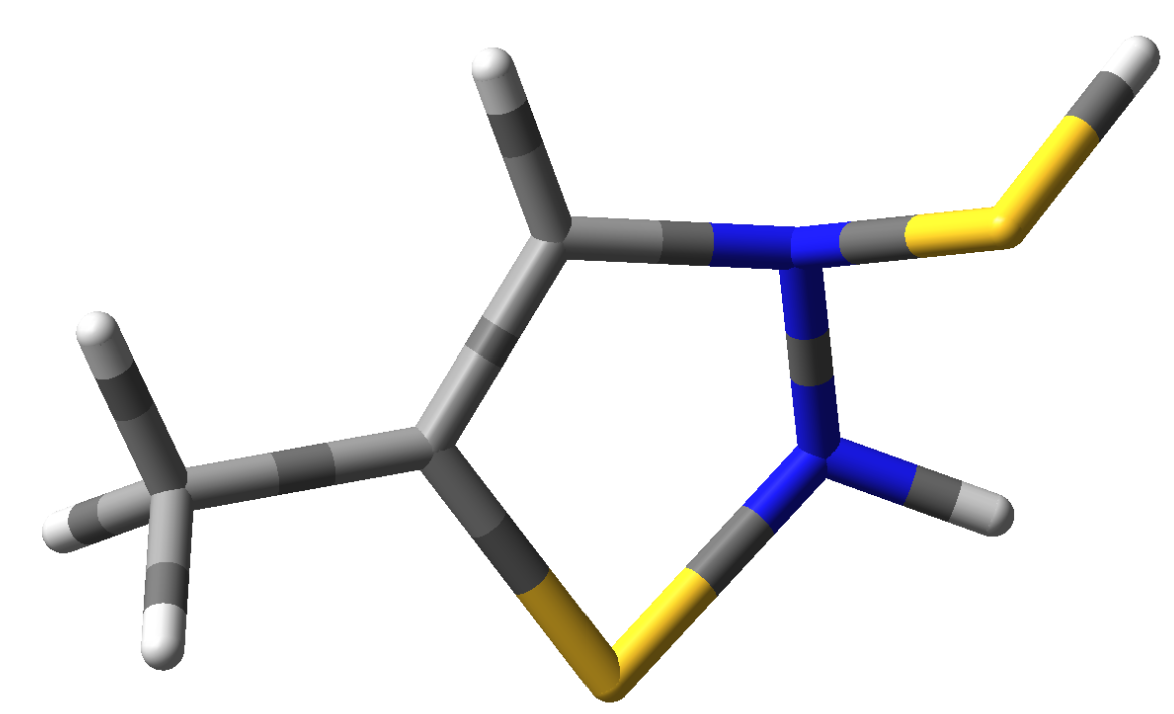} &
\includegraphics[width=0.125\textwidth]{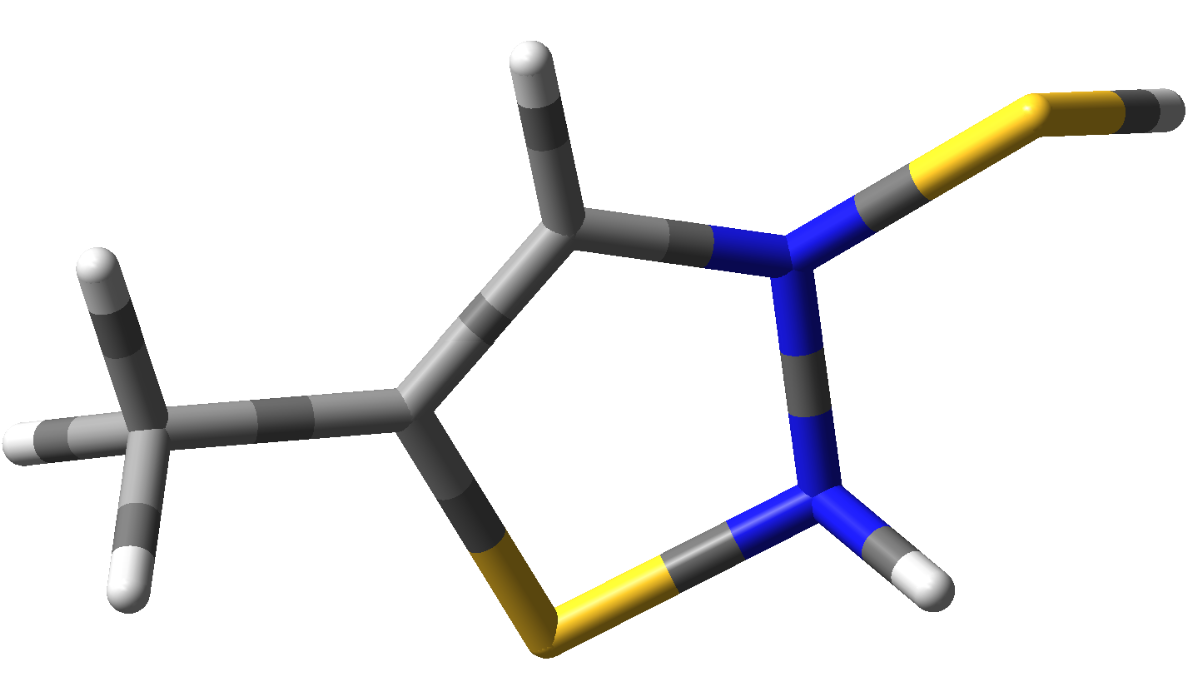} &
\includegraphics[width=0.115\textwidth]{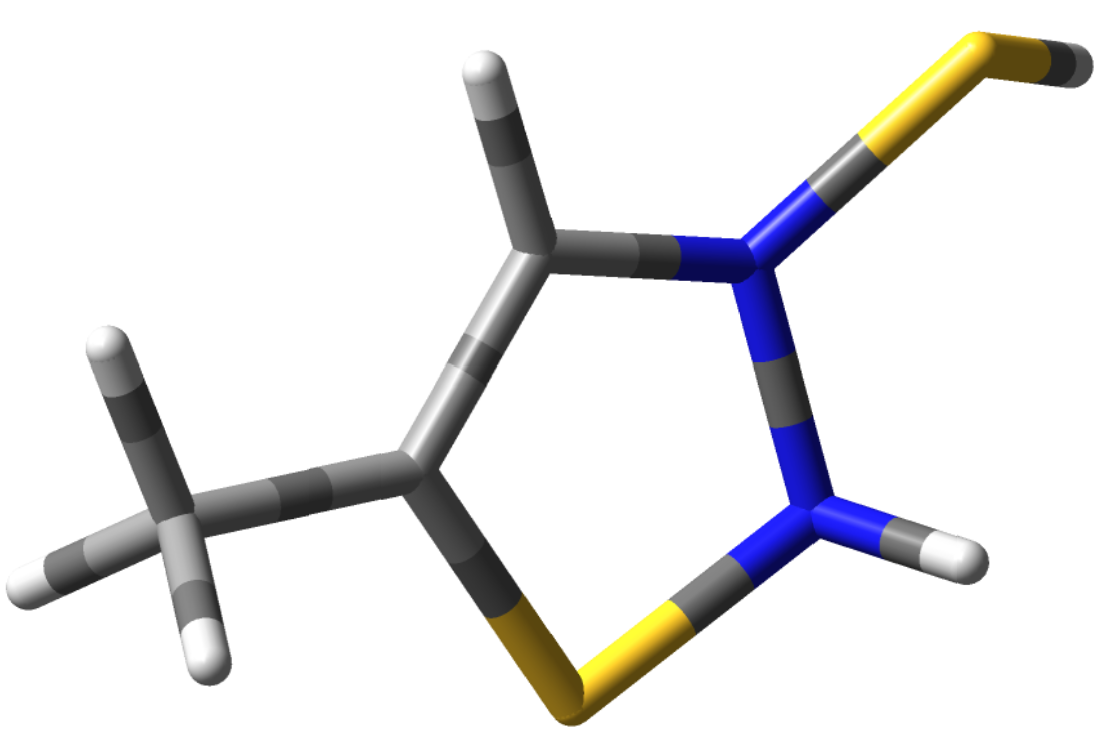} \\[-2pt]

RMSD (\AA) &
\textbf{0.0001} & 0.5718 & 1.1529 & 0.974 & 0.983 \\[-2pt]

$\Delta E$ (kcal/mol) &
\textbf{0.002} & 0.625 & 5.894 & 2.461 & 2.466 \\

\addlinespace[6pt]

\includegraphics[width=0.110\textwidth]{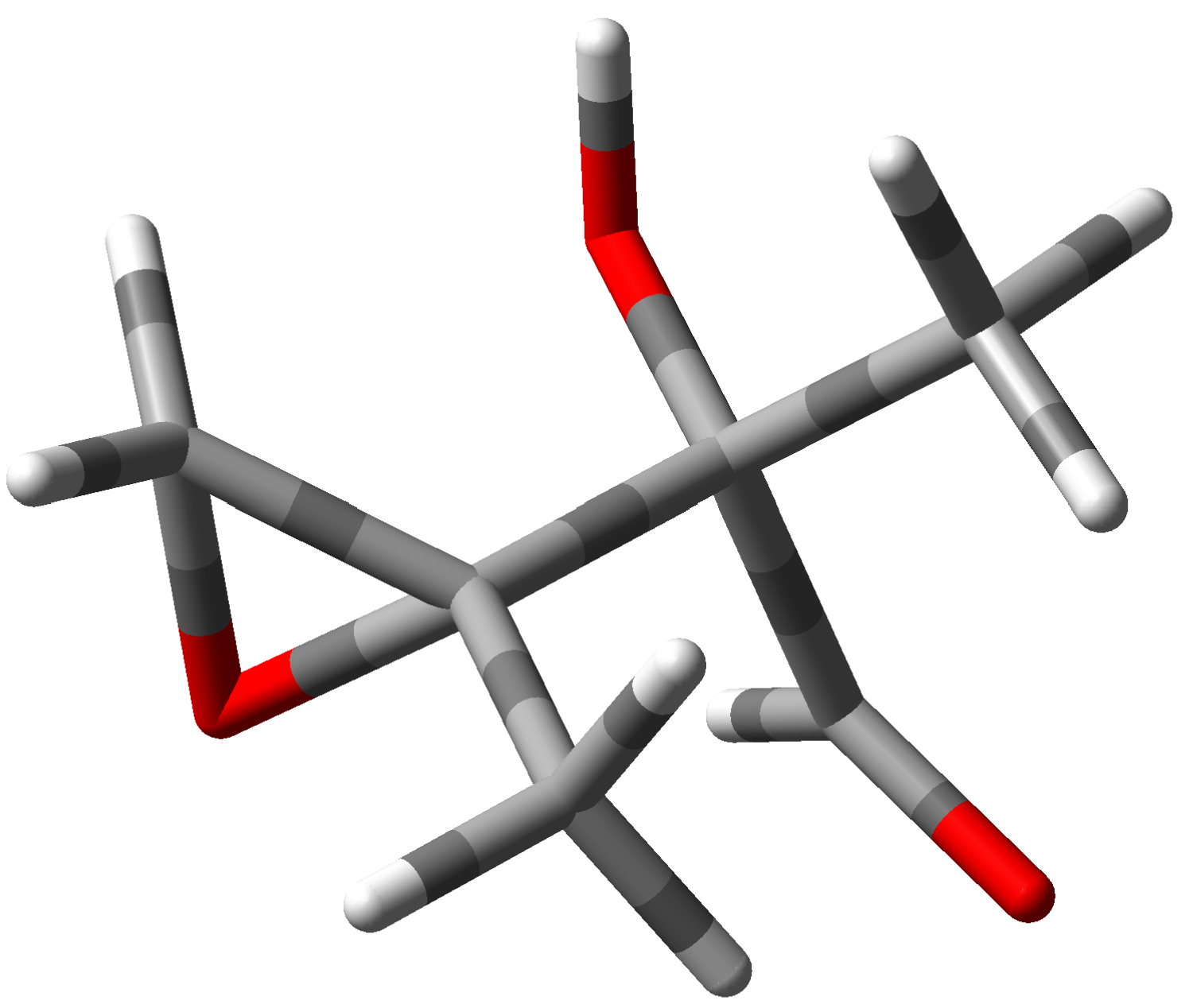} &
\includegraphics[width=0.115\textwidth]{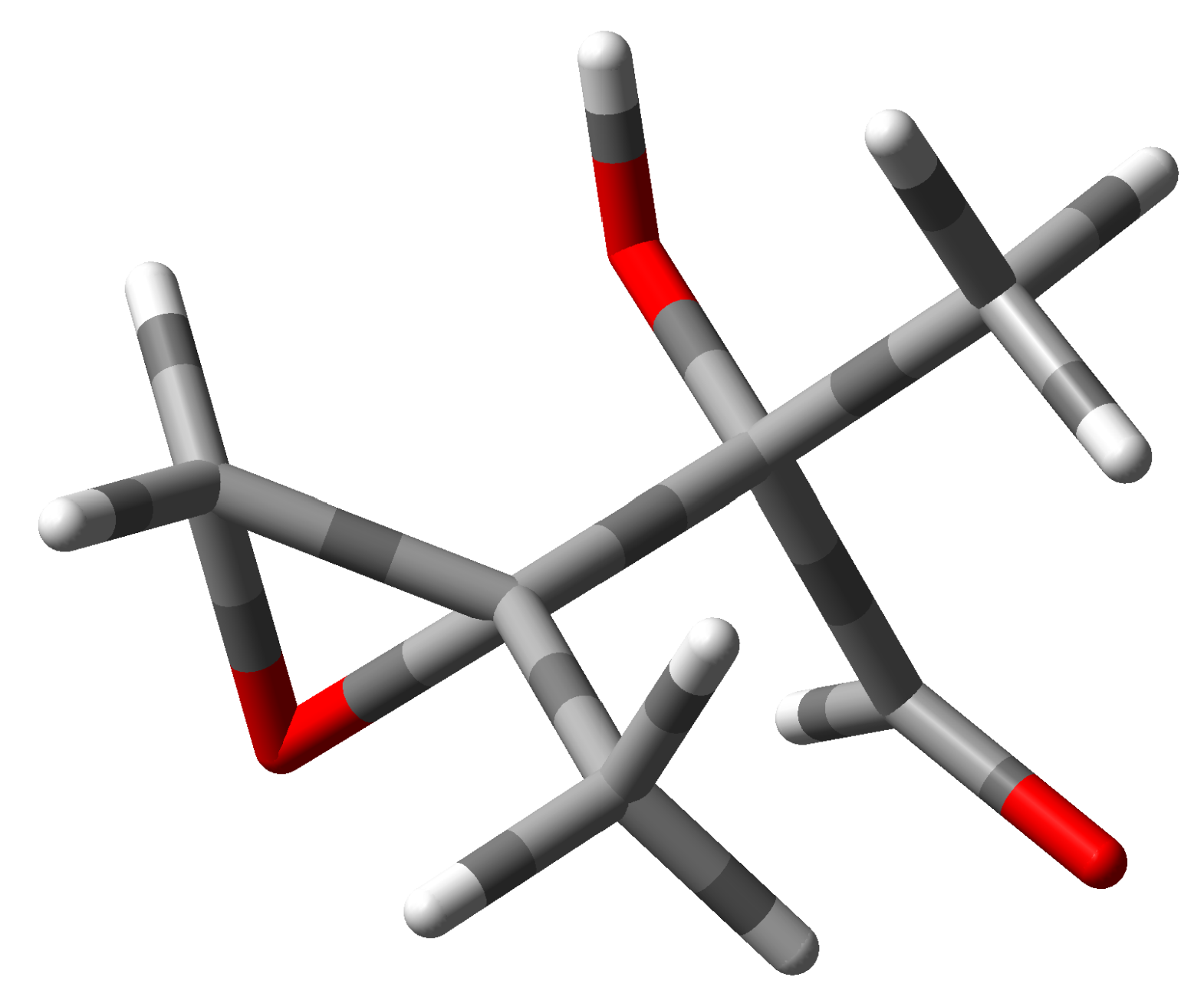} &
\includegraphics[width=0.086\textwidth]{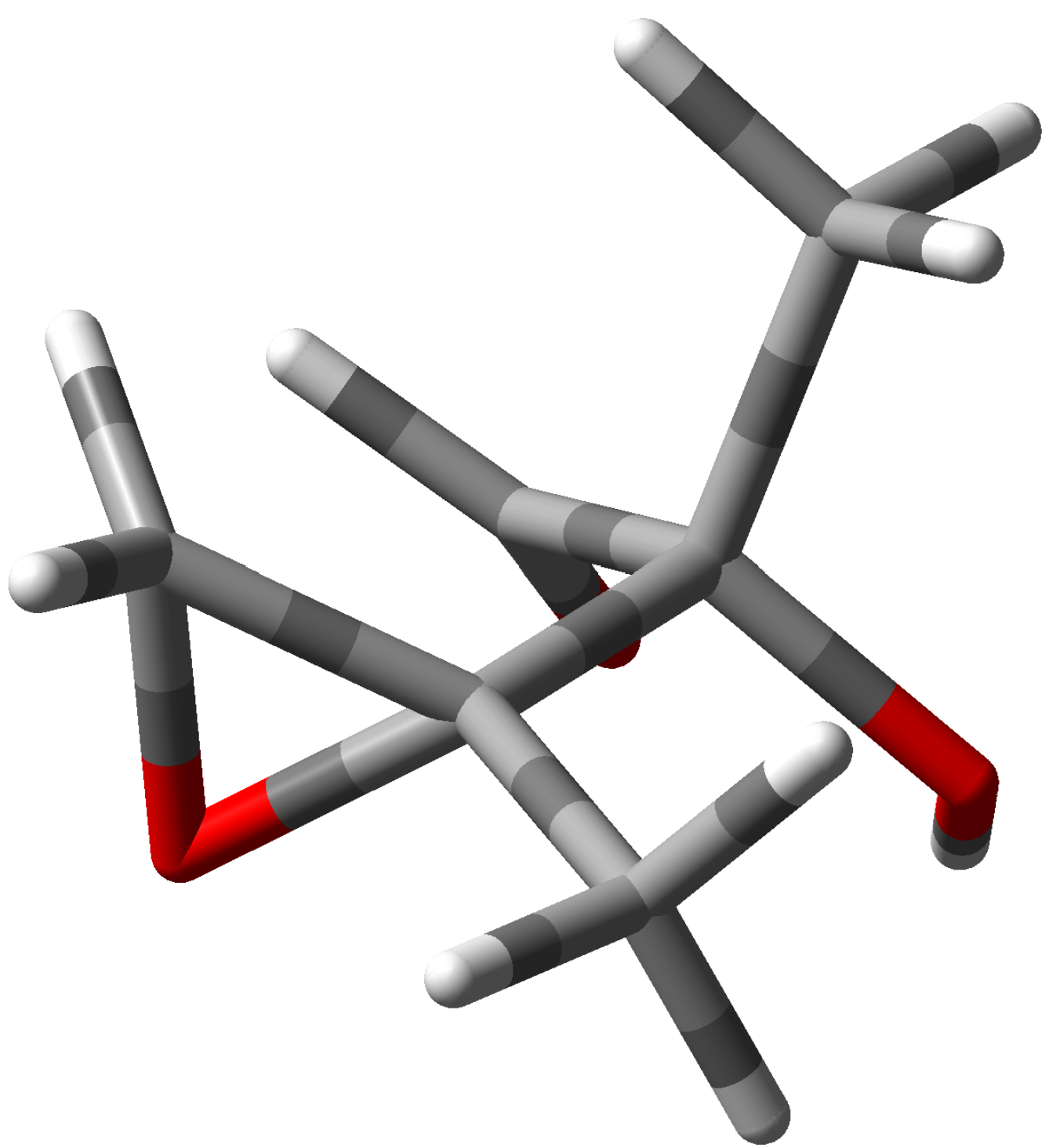} &
\includegraphics[width=0.110\textwidth]{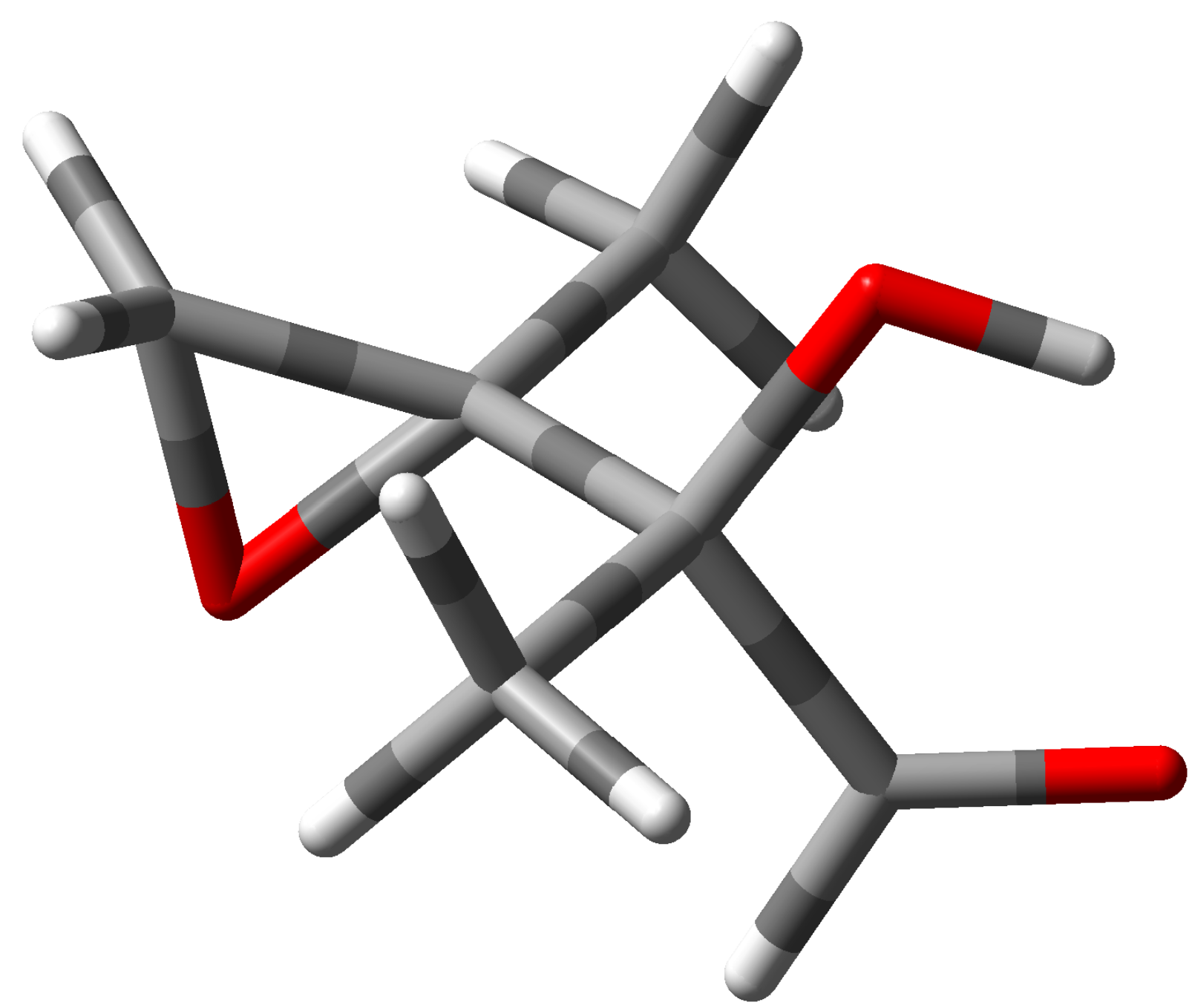} &
\includegraphics[width=0.110\textwidth]{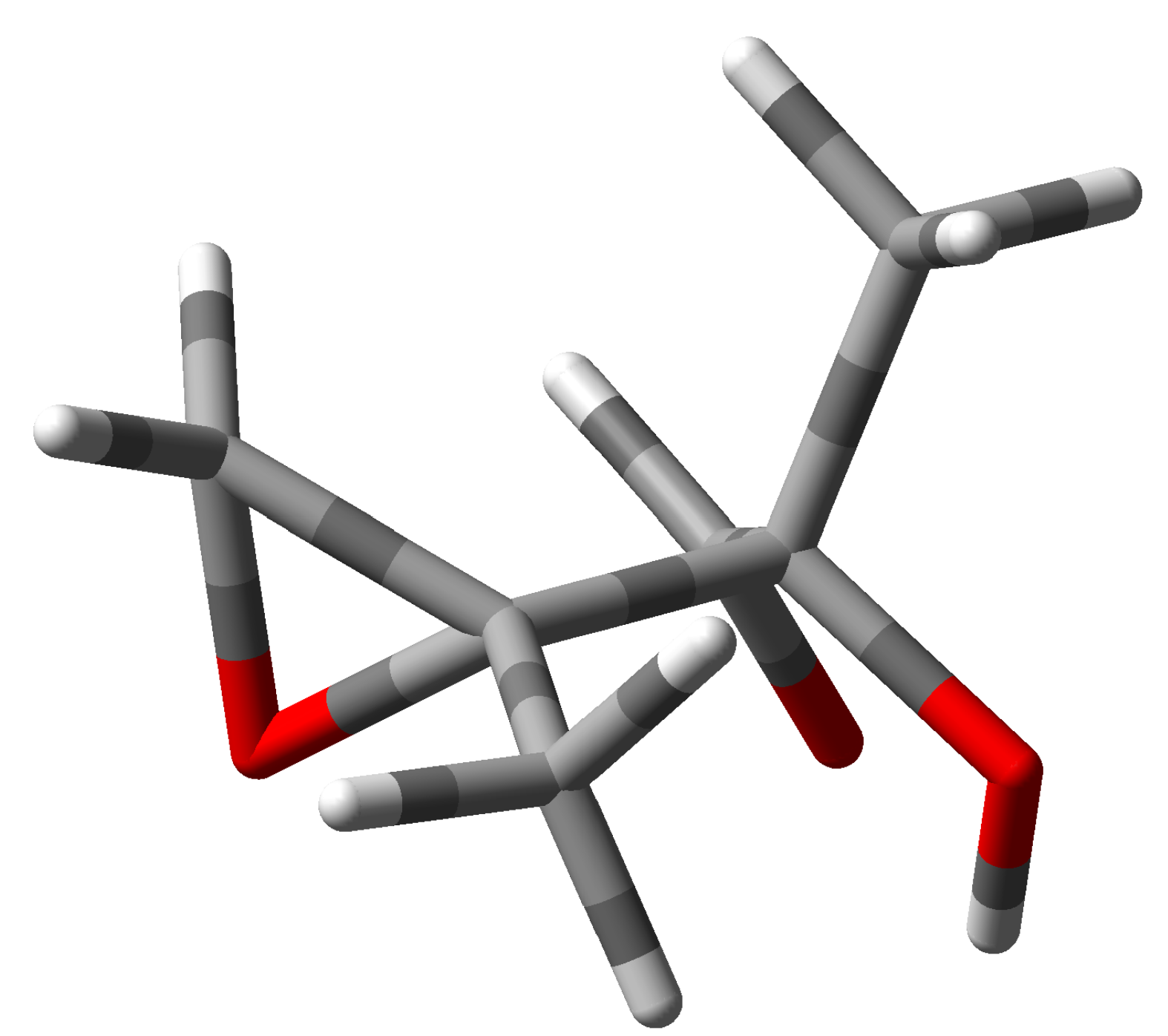} &
\includegraphics[width=0.105\textwidth]{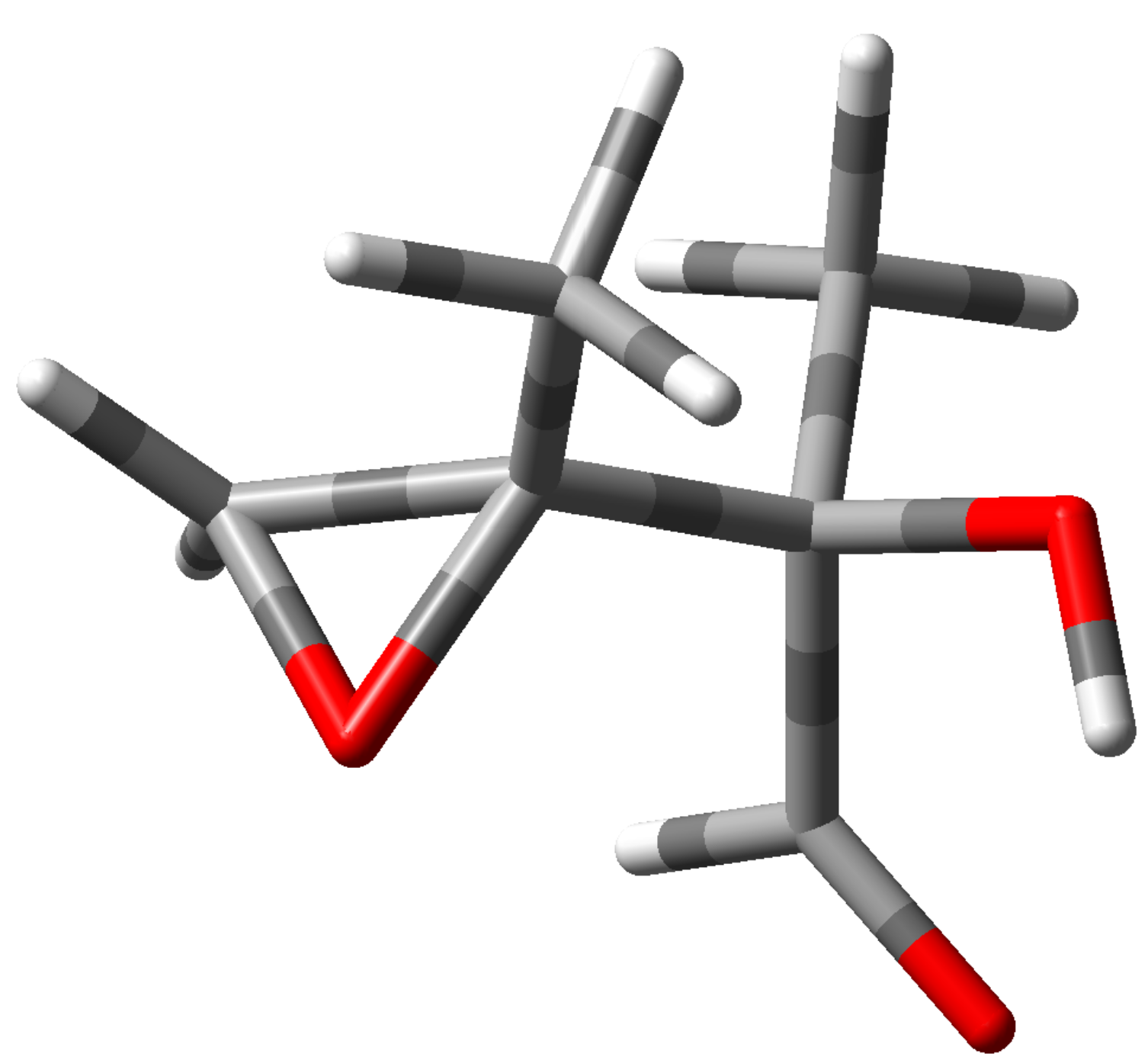} \\[-2pt]

RMSD (\AA) &
\textbf{0.0002} & 1.0337 & 0.8221 & 1.041 & 1.055 \\[-2pt]

$\Delta E$ (kcal/mol) &
\textbf{0.003} & 3.055 & 2.880 & 3.162 & 3.180 \\

\addlinespace[6pt]

\includegraphics[width=0.135\textwidth]{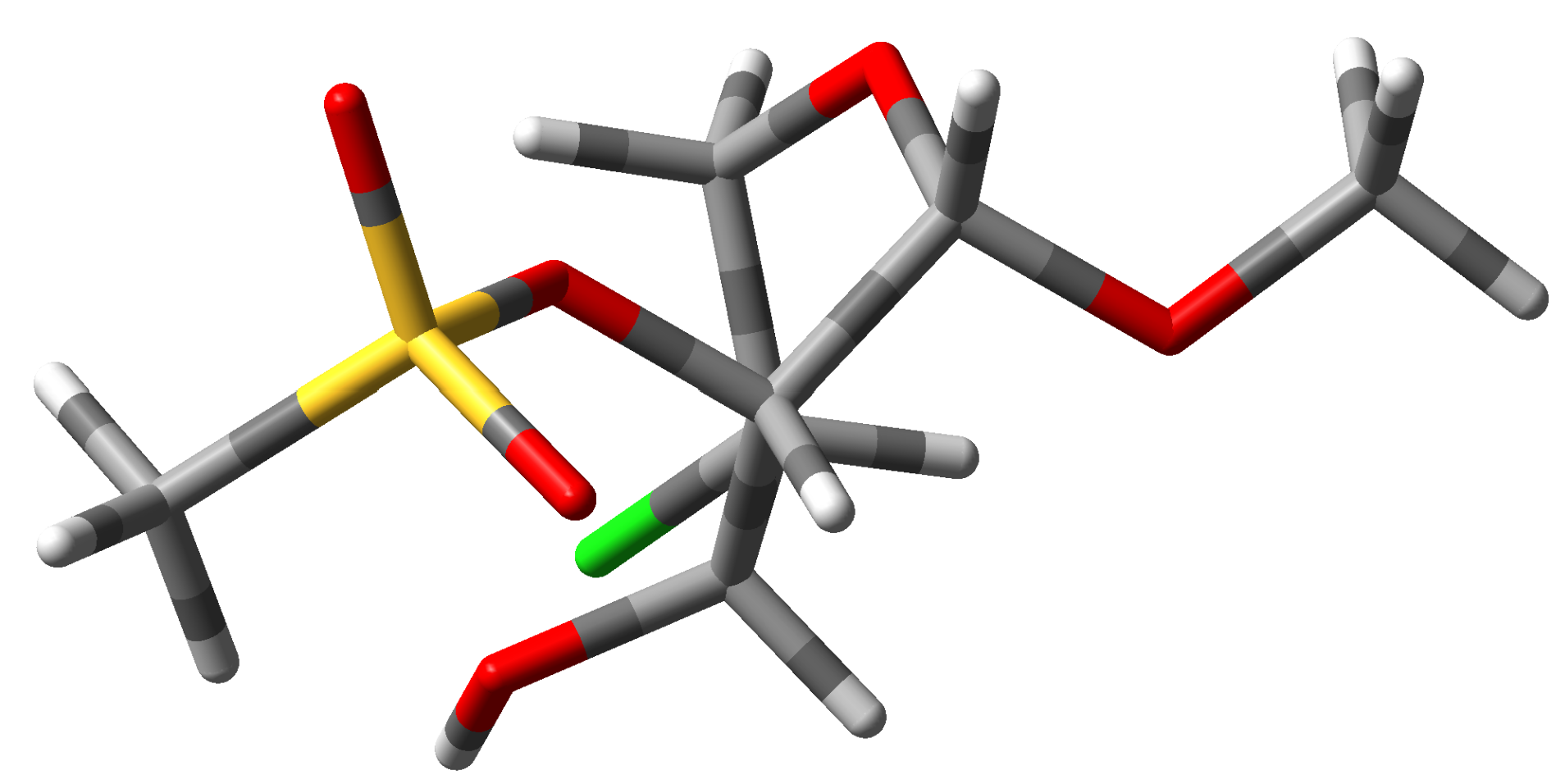} &
\includegraphics[width=0.135\textwidth]{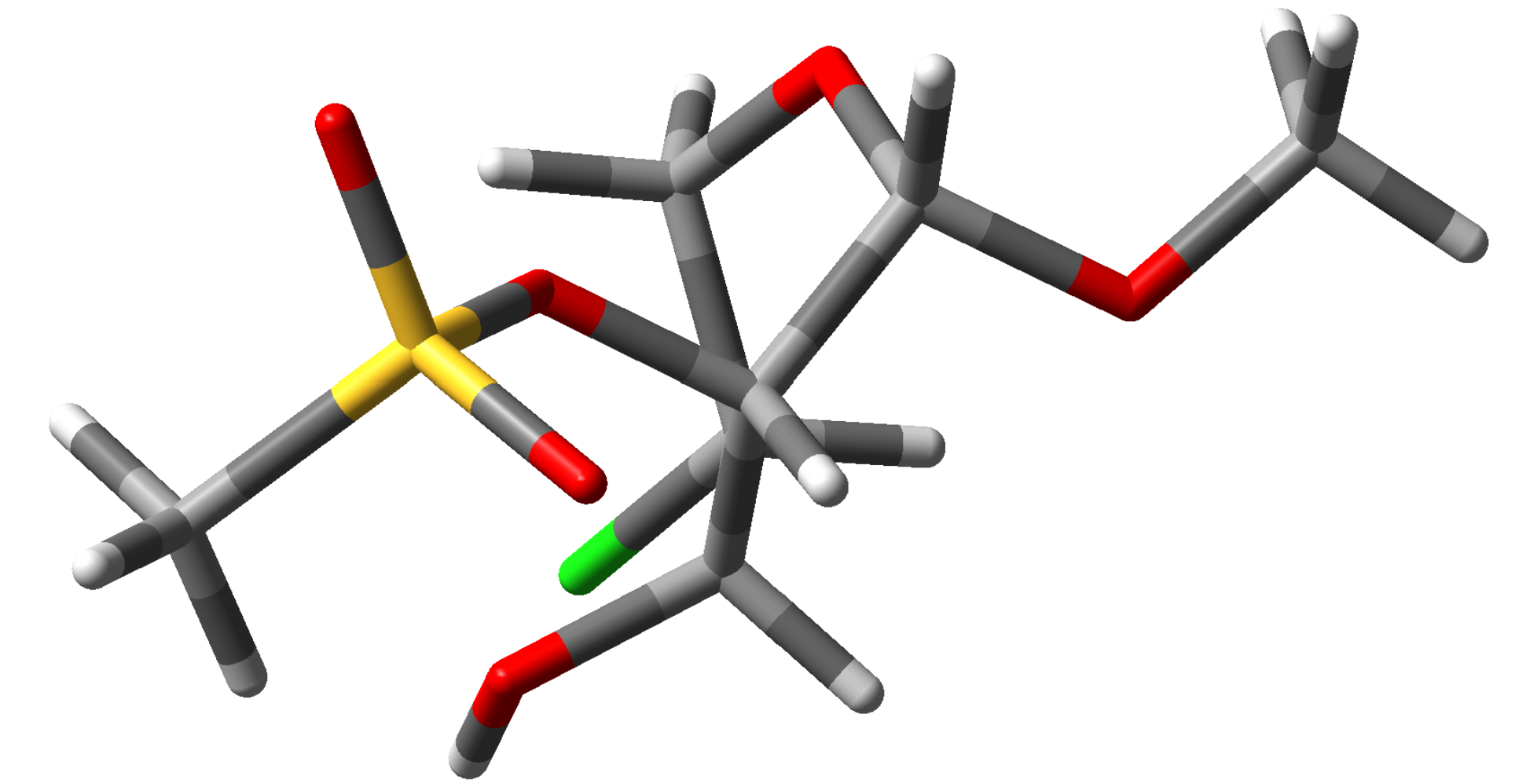} &
\includegraphics[width=0.130\textwidth]{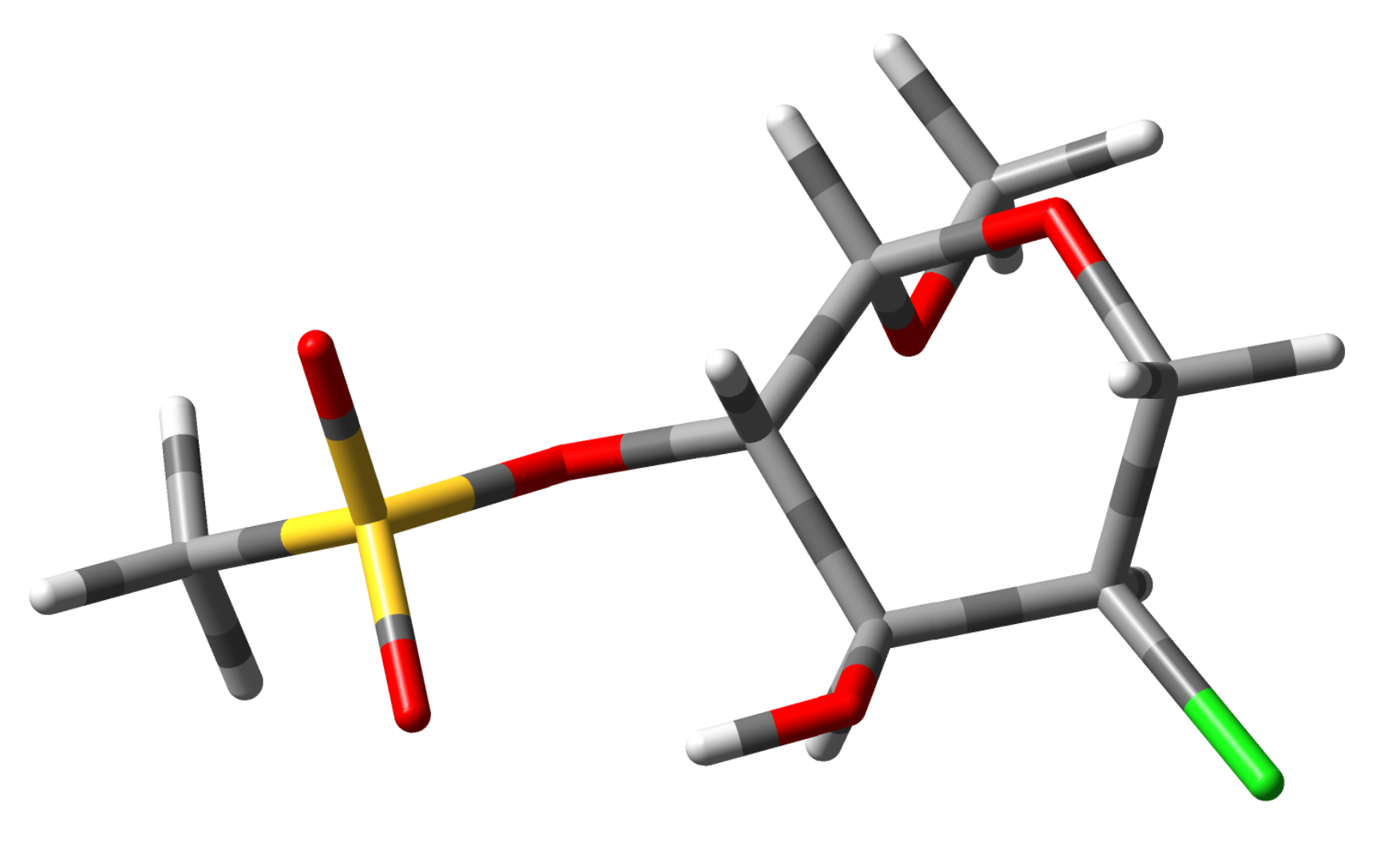} &
\includegraphics[width=0.110\textwidth]{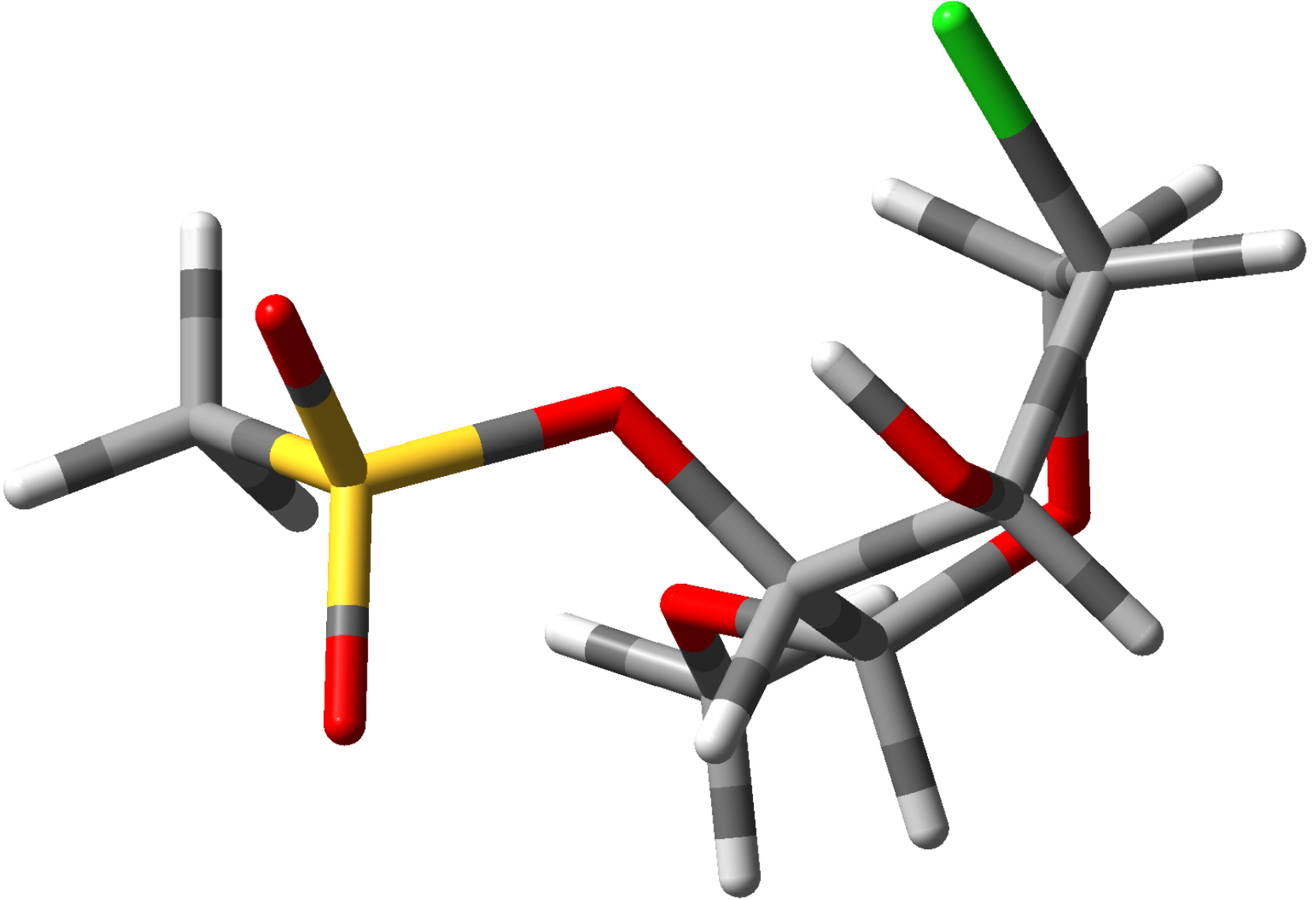} &
\includegraphics[width=0.115\textwidth]{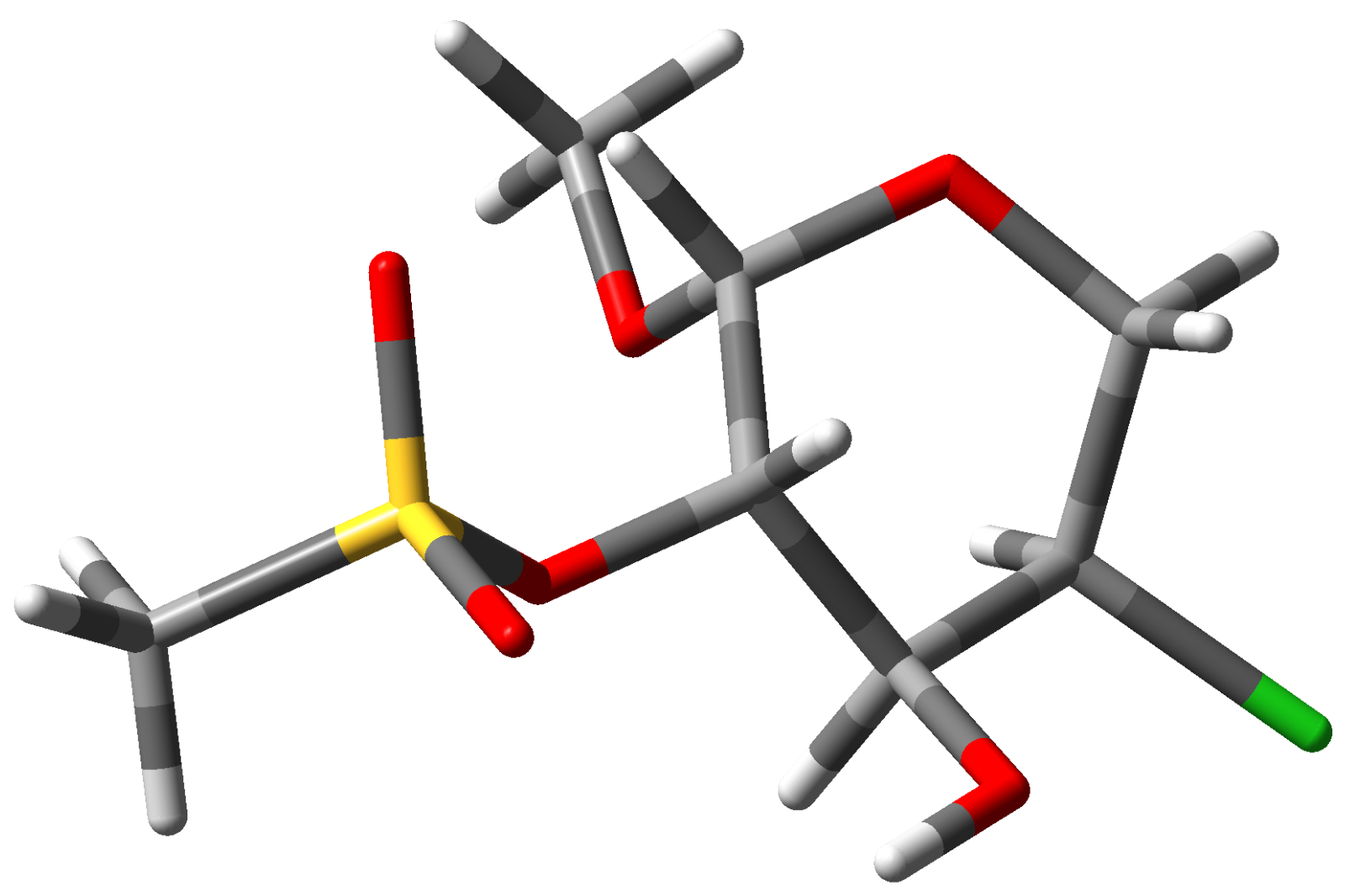} &
\includegraphics[width=0.115\textwidth]{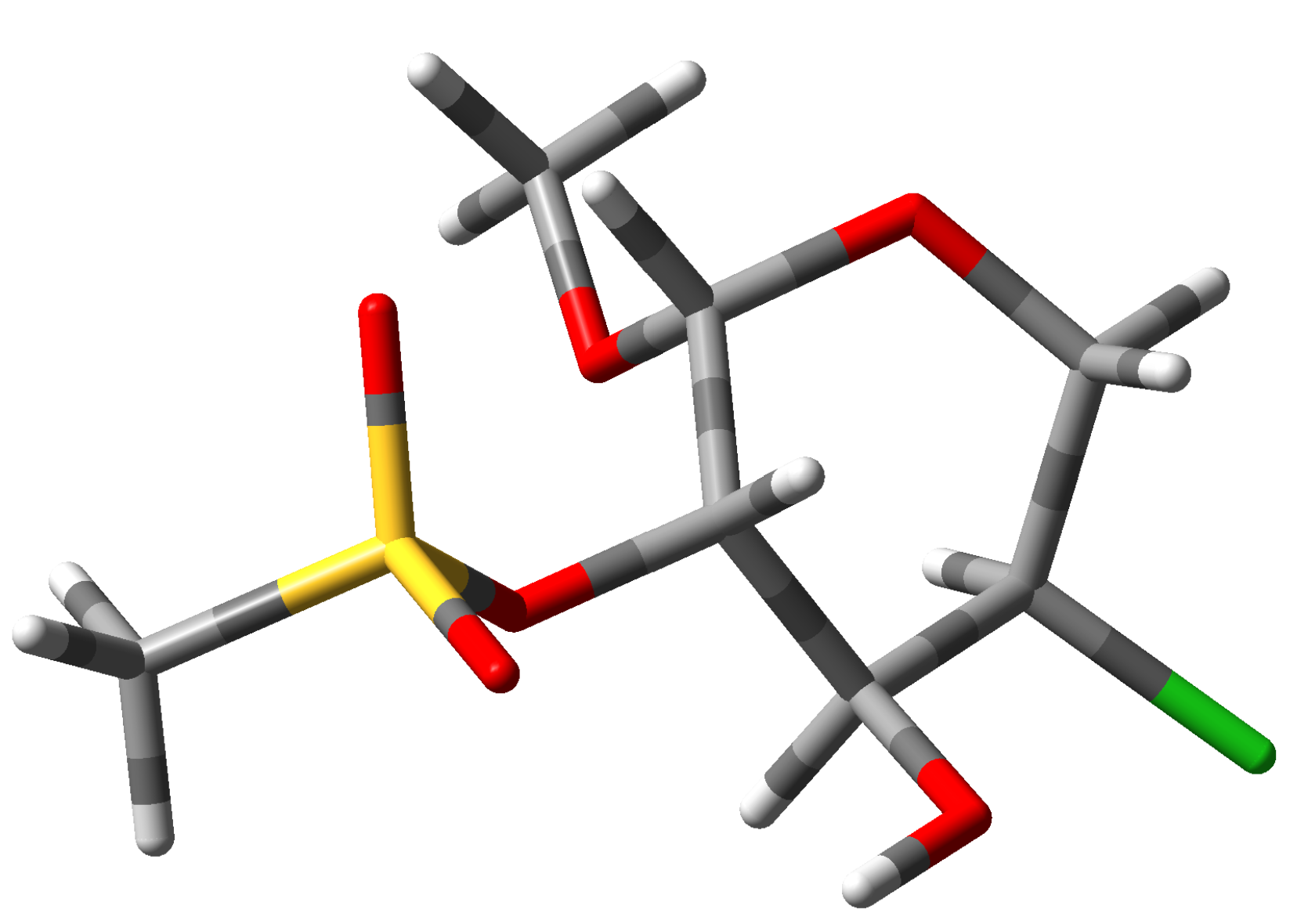} \\[-2pt]

RMSD (\AA) &
\textbf{0.0016} & 1.0774 & 1.2756 & 1.385 & 1.388 \\[-2pt]

$\Delta E$ (kcal/mol) &
\textbf{0.045} & 4.108 & 6.014 & 6.114 & 6.115 \\

\bottomrule
\end{tabular}
\end{table*}

\subsection{Reference Geometry Computations}
Initial 3D structures for all molecules were generated via RDKit’s ETKDG algorithm~\cite{riniker2015better} and pre-optimized with the MMFF94 force field~\cite{halgren1996merck}.  
\begin{itemize} 
  \item \textbf{External ``ZINC20'' test sets:} Final reference geometries and single-point energies were computed at the B3LYP/TZVP level using Gaussian 16 software, with default convergence settings.
\end{itemize}

\subsection{Data Preprocessing}
Coordinates were centered and standardized (zero mean, unit variance per axis) before model input. Atom types were one-hot encoded; bond adjacency and angular/dihedral graphs were constructed via RDKit v2024.03.5. Dataset splits follow an 80:10:10 train/val/test ratio stratified by heavy-atom count; external sets are used only for final evaluation.

\section{Results and discussion}

\subsection{Training protocol and model architecture}

GeoOpt-Net is a single-pass, SE(3)-equivariant geometry refinement model trained under a two-stage, multi-fidelity protocol (Fig.~\ref{fig2}).
The framework integrates transferable geometric pre-training with equivariant coordinate decoding to achieve target-level quantum-chemical accuracy in a single forward pass.

\paragraph{Two-stage multi-fidelity training.}
As illustrated in Fig.~\ref{fig2}a, GeoOpt-Net is trained using a two-stage strategy across quantum-chemical fidelity levels.
In Stage~1, the network is pre-trained on large-scale datasets computed at a lower level of theory (B3LYP/6-31G(2df,p)) to learn transferable geometric priors that capture generic bonding patterns, angular preferences, and torsional statistics.
In Stage~2, the pretrained weights are warm-started and fine-tuned on target-level B3LYP/TZVP data, enabling systematic correction of element- and basis-set-dependent deviations while preserving the global geometric structure learned during pre-training.

To condition geometric refinement on the target theoretical level, we introduce a fidelity-aware feature modulation (FAFM) mechanism,
\begin{equation}
\tilde{\mathbf{h}} = \mathbf{h} \odot (1 + \mathbf{g}_d) + \mathbf{b}_d,
\end{equation}
where $\mathbf{g}_d$ and $\mathbf{b}_d$ are learned scale and shift parameters derived from a domain embedding $d$.
This mechanism enables theory-specific geometric calibration without retraining the entire network, facilitating efficient transfer across quantum-chemical levels.

\paragraph{SE(3)-equivariant geometric architecture.}
The SE(3)-equivariant model architecture is shown in Fig.~\ref{fig2}b.
Starting from an RDKit ETKDG v3 + MMFF94 conformer, atomic Cartesian coordinates are projected into a 256-dimensional latent space using a linear embedding followed by LayerNorm and GELU activation~\cite{hendrycks2016gaussian}.

Molecular geometry is decomposed into three complementary equivariant streams corresponding to pairwise distances $r_{ij}$, three-body bond angles $\theta_{ijk}$, and four-body dihedral torsions $\varphi_{ijkl}$.
Scalar geometric features ($\ell=0$) are encoded via radial basis expansions $\phi(r_{ij})$, while directional information ($\ell\ge1$) is represented using real spherical harmonics $Y^{(\ell)}(\hat{\mathbf{r}}_{ij})$.

Equivariant message passing is performed via Clebsch--Gordan tensor products,
\begin{equation}
\mathbf{m}_{ij}^{(\ell)} =
\big( \mathbf{h}_i^{(\ell_1)} \otimes Y^{(\ell_2)}(\hat{\mathbf{r}}_{ij}) \big)_{\mathrm{CG}}
\cdot \phi(r_{ij}),
\end{equation}
which guarantees exact SE(3) equivariance by construction.
Nonlinear activations are applied exclusively to scalar channels, whereas vector channels undergo only learned linear updates and gated residual additions, following established equivariant architectures~\cite{fuchs2020se,schutt2021equivariant}.

The fused equivariant representation is decoded by a lightweight Transformer-based module to predict an equivariant coordinate refinement,
\begin{equation}
\mathbf{R}_{\mathrm{refined}} =
\mathbf{R}_{\mathrm{initial}} +
\mathcal{F}_{\theta}(\mathcal{G}, \mathbf{R}_{\mathrm{initial}}, d),
\end{equation}
yielding refined atomic coordinates that are invariant to global translations and equivariant under rotations.

\paragraph{Equivariance verification.}
To empirically validate SE(3) equivariance, we applied random global rotations and translations to 1{,}000 test structures and compared the predicted coordinates under
``rotate--then--predict'' and ``predict--then--rotate'' protocols.
The maximum deviation between the two procedures remained below $10^{-5}$\,\AA, confirming numerical SE(3) equivariance within floating-point precision (see~Fig.~S5).

\paragraph{Training objective.}
Model optimization minimizes a composite geometry-aware loss that jointly constrains global and local structural accuracy.
Specifically, the loss includes a global RMSD penalty on Cartesian coordinates ($L_{\mathrm{rmsd}}$), mean-squared-error losses on bond lengths, bond angles, and dihedral torsions ($L_{\mathrm{bond}}$, $L_{\mathrm{angle}}$, $L_{\mathrm{dihedral}}$), as well as a soft bond-length range constraint ($L_{\mathrm{bond\_range}}$) to suppress unphysical distortions.
All loss definitions, weights, and ablation analyses are provided in Tables~S1--S3.

Training is performed using AdamW (initial learning rate $1\times10^{-3}$, weight decay $1\times10^{-5}$), with learning-rate decay at epochs 50 and 75 and gradient clipping on mini-batches of 64 molecules.
This training protocol achieves sub-milli-\AA{} accuracy in a single forward pass while preserving strict SE(3) equivariance.

Such a multi-branch geometric design is critical.
Bond-only models neglect angular and torsional constraints, angle-only models miss long-range torsional flexibility, and point-cloud approaches treat atoms as orderless, discarding chemical connectivity~\cite{shi2021learning,zhu2022direct,xu2022geodiff,ganea2021geomol,jing2022torsional}.
In contrast, GeoOpt-Net explicitly models pairwise, three-body, and four-body geometric interactions, ensuring that each atomic representation encodes its full local coordination environment and thereby enhancing robustness and transferability across diverse chemical families.

\subsection{Geometric and Energetic Fidelity Relative to DFT References}

We benchmark the performance of \textbf{GeoOpt-Net} against representative structure generation and refinement methods, including UMA, xTB, Auto3D, and RDKit, using density functional theory (DFT) optimized geometries at the B3LYP/TZVP level as references. Figure~\ref{fig:geo_benchmark} summarizes qualitative comparisons and large-scale statistical analyses of geometric and energetic accuracy.

As illustrated by the representative examples summarized in Table~\ref{tab:three_example_compact}, GeoOpt-Net closely reproduces the B3LYP/TZVP reference geometries, yielding sub-milliångström RMSD values and negligible single-point energy deviations. In contrast, all baseline methods exhibit noticeable structural distortions, particularly in torsional arrangements and heteroatom-containing motifs, which lead to substantially larger RMSD values and increased energy errors. These discrepancies arise from local geometric inaccuracies rather than rigid-body misalignment, highlighting limitations in capturing fine-grained molecular structure.

Figure~\ref{fig:geo_benchmark}a presents the logarithmic distribution of all-atom RMSD values over the full benchmark set. GeoOpt-Net produces a sharply peaked distribution centered at $\log(\mathrm{RMSD}) \approx -4$, corresponding to sub-0.001~\AA\ accuracy for the majority of molecules. By comparison, UMA, xTB, Auto3D, and RDKit exhibit broader distributions centered between $\log(\mathrm{RMSD}) \approx -1$ and $0$, with extended tails indicating frequent large geometric deviations.

Consistent behavior is observed for single-point energy deviations at B3LYP/TZVP level of theory, shown in Fig.~\ref{fig:geo_benchmark}b. GeoOpt-Net yields a narrow energy distribution concentrated near zero, whereas baseline methods display broader and systematically shifted distributions, with typical errors on the order of several kcal~mol$^{-1}$. This strong correspondence between low RMSD and low $\Delta E$ confirms that the energetic fidelity of GeoOpt-Net arises from accurate geometries rather than explicit energy fitting (correlation analysis see~Fig.~S6).

To further elucidate the origin of these improvements, Fig.~\ref{fig:geo_benchmark}c--e decompose the geometric errors into bond length, bond angle, and dihedral angle components. GeoOpt-Net achieves near-numerical-precision bond lengths, sub-degree bond angle errors, and dihedral angle errors reduced by one to two orders of magnitude relative to all baseline methods. The pronounced reduction in torsional errors is particularly significant, as dihedral distortions are a dominant source of conformational and energetic inaccuracies in flexible molecules.

\begin{figure*}[bp]
  \centering
  \includegraphics[width=1\textwidth]{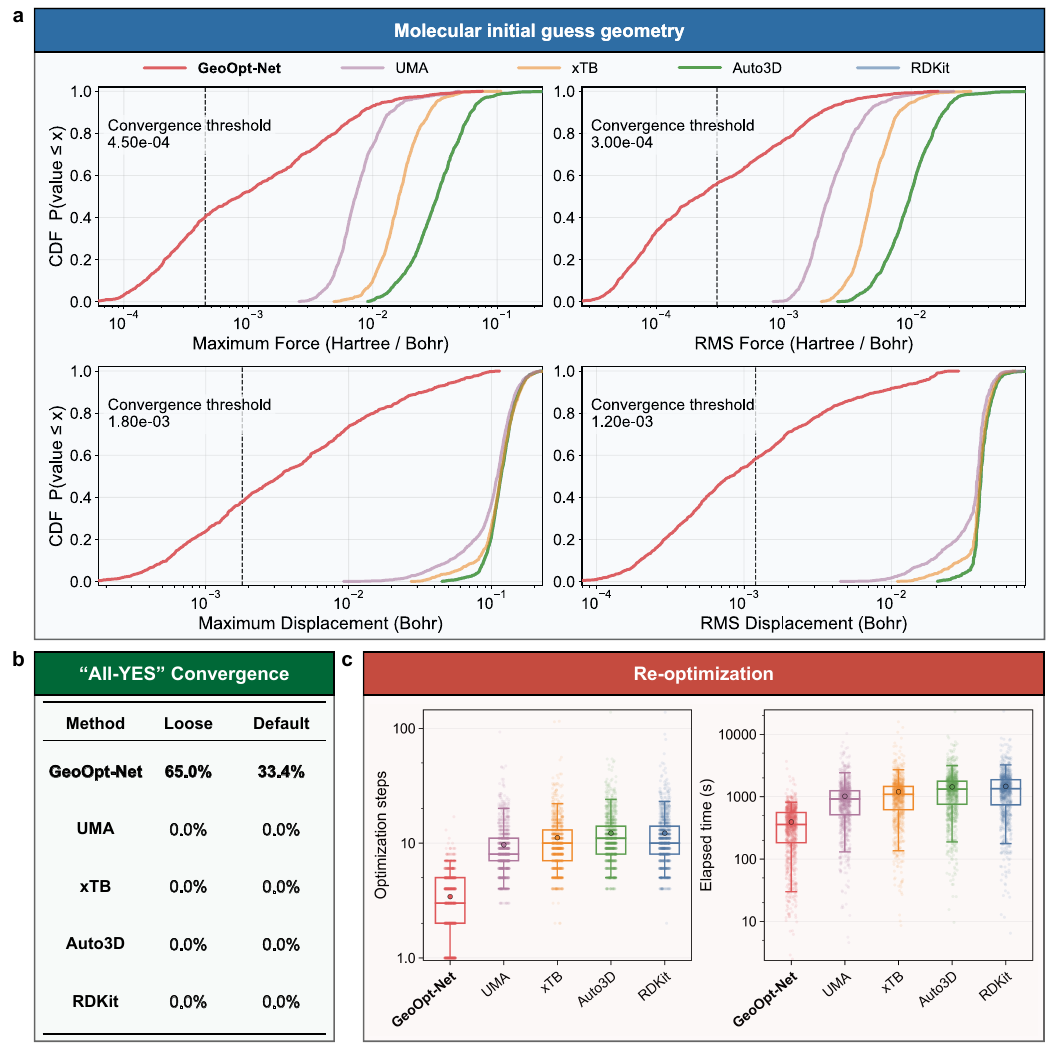}
    \caption{
    \textbf{Quality of molecular initial guess geometries and impact on DFT convergence.}
    (a) Cumulative distribution functions (CDFs) of maximum force, RMS force, maximum
    displacement, and RMS displacement for initial geometries generated by different methods.
    Dashed lines indicate the corresponding DFT geometry optimization convergence thresholds.
    (b) ``All-YES'' convergence rates, defined as the fraction of molecules simultaneously
    satisfying all four convergence criteria under loose and default thresholds.
    (c) DFT re-optimization performance starting from different initial geometries, quantified
    by the number of optimization steps (left) and total elapsed wall-clock time (right).
    }
    \label{fig:init_guess}
\end{figure*}

Overall, these results demonstrate that GeoOpt-Net enables B3LYP/TZVP-level geometric accuracy in a single forward pass, outperforming conventional force-field, semiempirical, and machine-learning-potential-based pipelines, and offering a scalable alternative for high-throughput quantum chemistry workflows.

\begin{figure*}[tbp]
  \centering
  \includegraphics[width=0.8\textwidth]{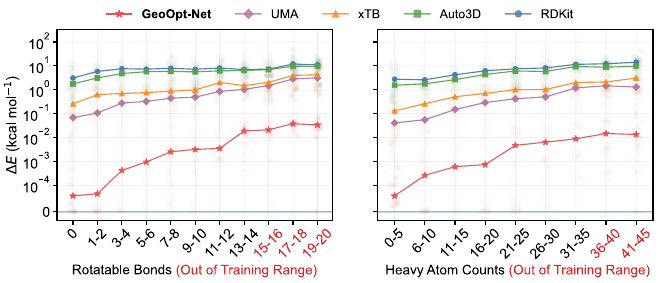}
    \caption{
    \textbf{Energy scaling with molecular complexity.}
    DFT single-point energy deviations ($\Delta E$, B3LYP/TZVP) evaluated as a function of
    molecular complexity, quantified by the number of rotatable bonds (left) and heavy atom
    counts (right), for different molecular geometry initialization approaches.
    }
    \label{fig:robustness_scaling}
\end{figure*}

\subsection{Impact of Initial Geometry Quality on DFT Convergence Efficiency}

We evaluate the quality of molecular initial guess geometries generated by GeoOpt-Net and baseline methods (UMA, xTB, Auto3D, and RDKit) by directly assessing their proximity to standard density functional theory (DFT) geometry optimization convergence criteria at the B3LYP/TZVP level of theory. Specifically, we examine four key quantities commonly used to determine convergence during geometry optimization: maximum force, RMS force, maximum displacement, and RMS displacement.

As shown in Fig.~\ref{fig:init_guess}a, GeoOpt-Net consistently produces initial geometries that are substantially closer to the DFT convergence thresholds than all baseline methods across all four metrics. The cumulative distribution functions (CDFs) indicate that a significant fraction of GeoOpt-Net structures already satisfy, or nearly satisfy, the corresponding force and displacement criteria. At the B3LYP/TZVP level, approximately 40.1\% and 56.0\% of GeoOpt-Net geometries satisfy the maximum force and RMS force criteria, respectively, while 37.8\% and 58.3\% meet the maximum and RMS displacement thresholds. In contrast, the distributions associated with UMA, xTB, Auto3D, and RDKit remain far from the convergence thresholds, with negligible probability mass in the near-converged regime.

This separation directly translates into practical convergence behavior. Under the commonly used ``All-YES'' criterion, which requires all four convergence conditions to be simultaneously satisfied, GeoOpt-Net achieves nonzero convergence rates of 65.0\% under loose thresholds and 33.4\% under default thresholds (Fig.~\ref{fig:init_guess}b). In contrast, none of the baseline methods produce structures that satisfy all convergence criteria, indicating a qualitative difference in the suitability of the initial geometries for DFT optimization.

To further quantify the downstream impact, we perform full DFT re-optimization starting from the initial geometries generated by each method (Fig.~\ref{fig:init_guess}c). Initial geometries provided by GeoOpt-Net require substantially fewer optimization steps to reach convergence and consistently reduce the total wall-clock time. This improvement reflects the fact that GeoOpt-Net places molecular structures much closer to the region defined by the DFT convergence criteria, whereas baseline methods typically start far from this region.

Overall, these results demonstrate that GeoOpt-Net generates initial geometries that are intrinsically compatible with DFT convergence requirements at the B3LYP/TZVP level of theory, leading to higher convergence rates and significantly reduced computational cost in quantum chemical geometry optimization workflows.

\begin{figure*}[htbp]
  \centering
  \includegraphics[width=0.954\textwidth]{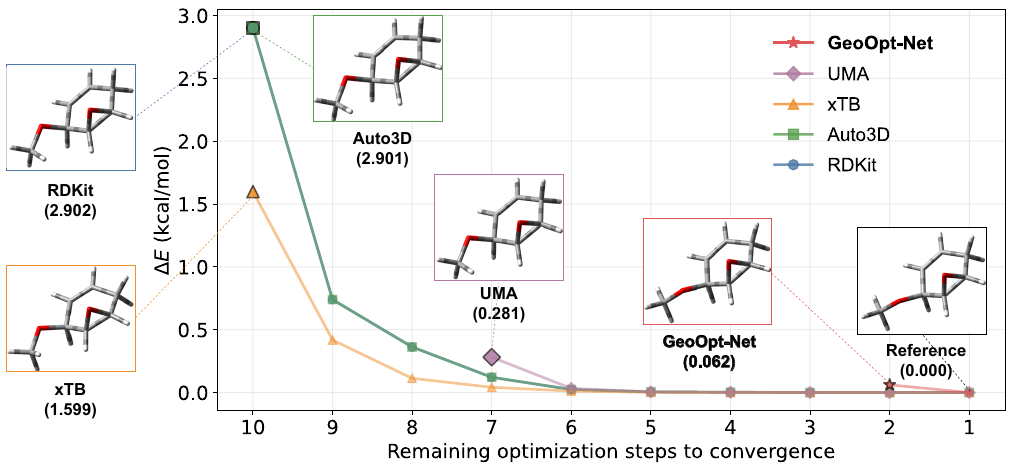}
    \caption{ \textbf{Geometric proximity and DFT convergence analysis .} DFT single-point energy deviation ($\Delta E$, B3LYP/TZVP) plotted against the remaining number of geometry-optimization steps required to reach convergence, illustrating how initial structures generated by different methods relate to the DFT convergence process. }
    \label{fig:robustness_optimization}
\end{figure*}

\subsection{Robustness, Scaling Behavior, and Physical Consistency beyond the Training Domain}

We further evaluate the robustness and physical reliability of GeoOpt-Net by analyzing how
DFT single-point energy deviations ($\Delta E$, B3LYP/TZVP) evolve with increasing molecular
complexity.
All analyses are performed on 1,000 molecules uniformly sampled from a billion-scale
drug-like molecular library, providing a chemically diverse and practically relevant
evaluation set that extends well beyond the molecular complexity covered during training.

Figure~\ref{fig:robustness_scaling} summarizes the scaling behavior of $\Delta E$ as a
function of molecular complexity, quantified by the number of rotatable bonds (Fig.~\ref{fig:robustness_scaling} left) and heavy atom counts (Fig.~\ref{fig:robustness_scaling} right).
Across both measures, all methods exhibit increasing energy deviations as the molecular
size and flexibility grow, reflecting the intrinsic difficulty of generating accurate
geometries for larger and more conformationally complex systems.
However, the magnitude and rate of error accumulation differ substantially among methods.

Conventional structure generators, including RDKit and Auto3D, consistently exhibit large
energy deviations already at low to moderate complexity, with typical errors on the order of 1--10~kcal~mol$^{-1}$ range that further increase for more complex molecules.
xTB and UMA partially mitigate these errors but still display a clear and systematic growth,
reaching the kcal~mol$^{-1}$ scale as either the number of rotatable bonds or heavy atom
counts increase.

In contrast, GeoOpt-Net maintains energy deviations that are orders of magnitude lower
across the entire complexity range.
For the vast majority of sampled molecules, $\Delta E$ remains well below
0.1~kcal~mol$^{-1}$, with a smooth and gradual increase as molecular complexity grows.
Notably, this behavior persists in the high-complexity regimes highlighted in
Fig.~\ref{fig:robustness_scaling}, which lies well beyond the training domain of GeoOpt-Net.
The absence of abrupt error escalation in these regions indicates robust extrapolative
behavior rather than reliance on interpolation within the training set.

Beyond static complexity measures, Fig.~\ref{fig:robustness_optimization} links the DFT single-point energy deviation ($\Delta E$) to the remaining number of geometry-optimization steps required to reach convergence, thereby providing a physically meaningful metric for assessing geometric proximity to DFT convergence.
Structures generated by RDKit, xTB, UMA, and Auto3D are typically initialized far from the DFT convergence region and therefore require substantial structural relaxation, as indicated by both large initial $\Delta E$ values and long remaining optimization trajectories.
By contrast, GeoOpt-Net consistently places molecular geometries much closer to DFT convergence, yielding an initial $\Delta E$ of only 0.062~kcal~mol$^{-1}$ and requiring markedly fewer optimization steps.
Beyond energetic accuracy, this geometric improvement translates directly into enhanced electronic-structure consistency, as evidenced by the accurate reproduction of the dipole moment $\Delta\mu$.
As shown in Table~\ref{tab:dipole_moment}, dipole moments computed from GeoOpt-Net geometries deviate by only 0.002~Debye from the DFT reference, whereas all baseline methods systematically underestimate the dipole moment by 0.37--0.50~Debye. This comparison demonstrates that reduced geometric deviation in GeoOpt-Net leads to substantially improved electronic observables.
Taken together, these results show that GeoOpt-Net provides geometries that are not only energetically closer to DFT convergence but also preserves key electronic-structure properties, highlighting the physical consistency of the method beyond its training domain.

\begin{table}[tbp]
\centering
\caption{Comparison of dipole moments computed from molecular geometries optimized by GeoOpt-Net and baseline methods.
All dipole moments ($\mu$) are evaluated at the B3LYP/TZVP level of theory using the corresponding optimized geometries.
$\Delta\mu$ denotes the deviation from the reference value obtained from the DFT-converged geometry at the B3LYP/TZVP level of theory.}
\label{tab:dipole_moment}
\begin{tabular}{lcc}
\toprule
Method & $\mu$ (Debye) & $\Delta\mu$ (Debye) \\
\midrule
Reference   & 3.165 & 0.000 \\
GeoOpt-Net  & \textbf{3.167} & \textbf{+0.002} \\
UMA         & 2.796 & $-$0.369 \\
xTB         & 2.751 & $-$0.414 \\
Auto3D      & 2.669 & $-$0.496 \\
RDKit       & 2.667 & $-$0.498 \\
\bottomrule
\end{tabular}
\end{table}

\section{Conclusion}
In this work, we introduced GeoOpt-Net, a multi-scale SE(3)-equivariant geometry refinement framework that bridges inexpensive conformer generation and high-level DFT accuracy at B3LYP/TZVP level of theory. From a methodological perspective, existing approaches for generating DFT-ready structures span several distinct paradigms, including classical conformer generators based on distance geometry and force fields, semiempirical quantum methods, data-driven geometry refinement pipelines, and machine-learning interatomic potentials. GeoOpt-Net belongs to the data-driven refinement category but differs fundamentally from existing approaches by explicitly decoupling bond lengths, bond angles, and dihedral torsions into dedicated equivariant graph streams, which are fused through a lightweight Transformer decoder to capture chemically meaningful local coordination and long-range torsional couplings. This architecture is further combined with a two-stage training strategy, in which broadly transferable geometric representations are first learned by large-scale pretraining and subsequently fine-tuned to higher levels of theory, enabling systematic correction of basis-set- and element-dependent geometric deviations.

Comprehensive benchmarks against RDKit~\cite{Landrum2016RDKit2016094}, xTB~\cite{bannwarth2019gfn2}, Auto3D~\cite{liu2022auto3d}, and UMA~\cite{wood2025family} using B3LYP/TZVP reference data demonstrate that GeoOpt-Net produces near-DFT-quality geometries in a single forward pass, achieving sub-milli\AA{} all-atom RMSD and tightly concentrated single-point energy deviations. Ablation experiments further confirm that explicit SE(3) equivariance and spectral graph encodings are indispensable for controlling error growth. Error decomposition reveals that the dominant performance gains arise from substantially improved torsional accuracy, with dihedral errors reduced by one to two orders of magnitude. This distinction directly reflects methodological differences: classical and semiempirical approaches are limited by approximate energy models, while potential-based methods rely on iterative relaxation, whereas GeoOpt-Net directly learns a theory-aware geometric mapping that explicitly resolves torsional correlations.

Beyond static geometric accuracy, GeoOpt-Net delivers direct and practical benefits for quantum-chemical workflows. Initial structures optimized by GeoOpt-Net lie markedly closer to standard DFT convergence criteria, resulting in nonzero ``All-YES'' convergence rates (65.0\% under loose and 33.4\% under default thresholds), whereas all baseline methods fail to satisfy these criteria. Consequently, subsequent DFT re-optimizations require fewer optimization steps and reduced wall-clock time, indicating that GeoOpt-Net provides initial geometries already close to DFT-converged structures rather than merely improving rigid-body alignment.

Importantly, GeoOpt-Net remains robust beyond its training distribution. On external drug-like molecules from the ZINC20 database with increased molecular size and conformational flexibility, energy deviations increase smoothly and monotonically with molecular complexity, while key electronic observables such as dipole moments are preserved with near-reference fidelity. Together, these results demonstrate that multi-fidelity fine-tuning enables GeoOpt-Net to retain transferable geometric priors while achieving theory-aware refinement, establishing it as a scalable, physically consistent, and computationally efficient framework for generating DFT-ready geometries in large-scale quantum-chemical workflows.

\section{ASSOCIATED CONTENT}

\subsection{Data Availability}

The computational models and data sets reported in this work will be made available on GitHub.

\subsection{Supporting Information}

Details on the construction of the dataset (including data extraction and dataset composition), experimental settings and training procedures, and additional experimental details.

\section{AUTHOR INFORMATION}

\subsection{Corresponding Author}

\textbf{Fanyang Mo -- }
{School of AI for Science, Peking University Shenzhen Graduate School, Shenzhen, 518055, China; AI for Science (AI4S)-Preferred Program, Peking University Shenzhen Graduate School, Shenzhen 518055, China; State Key Laboratory of Advanced Waterproof Materials, School of Materials Science and Engineering, Peking University, Beijing 100871, China; School of Advanced Materials, Peking University Shenzhen Graduate School, Shenzhen 518055, China; Guangdong Provincial Key Laboratory of Nano-Micro Materials Research, Peking University Shenzhen Graduate School, Shenzhen 518055, China}; 
ORCID: 0000-0002-4140-3020; 
Email: fmo@pku.edu.cn

\subsection{Author}
\textbf{Chengchun Liu -- }
{School of AI for Science, Peking University Shenzhen Graduate School, Shenzhen, 518055, China; AI for Science (AI4S)-Preferred Program, Peking University Shenzhen Graduate School, Shenzhen 518055, China; State Key Laboratory of Advanced Waterproof Materials, School of Materials Science and Engineering, Peking University, Beijing 100871, China}; 
ORCID: 0009-0002-5550-4145

\textbf{Wendi Cai -- }
{School of AI for Science, Peking University Shenzhen Graduate School, Shenzhen, 518055, China; AI for Science (AI4S)-Preferred Program, Peking University Shenzhen Graduate School, Shenzhen 518055, China; School of Advanced Materials, Peking University Shenzhen Graduate School, Shenzhen 518055, China};

\textbf{Boxuan Zhao -- }
{School of AI for Science, Peking University Shenzhen Graduate School, Shenzhen, 518055, China; AI for Science (AI4S)-Preferred Program, Peking University Shenzhen Graduate School, Shenzhen 518055, China; State Key Laboratory of Advanced Waterproof Materials, School of Materials Science and Engineering, Peking University, Beijing 100871, China}; 

\subsection{Author contributions}
F.M. and C.L. conceived the project, designed the method, and analyzed the results. C.L. organized and prepared structural data. W.C. and B.Z. participated in the discussion and provided key suggestions. F.M. and C.L. wrote the paper. F.M. supervised the project.

\subsection{Notes}

The authors declare no competing financial interest.

\section{ACKNOWLEDGMENTS}

We thank Peking University Shenzhen Graduate School and Shenzhen Government for the start-up funding support. We thank the High-Performance Computing Platform of Peking University for machine learning model training. We appreciate Dr. Jianning Zhang for his constructive suggestions and insightful discussions.

\bibliography{achemso-demo}

\end{document}